# Ultra-high Q-factor superconducting tantalum resonators on 300 mm Si wafers.

R. Acharya[1], D. Perez Lozano[1], Ts. Ivanov[1], S. Massar[1], C. Vrancken[1], Y. Canvel[1], Y. Li[1], A.M. Vadiraj[1], J. Van Damme[1], S. Aghaeimeibodi[2], A. Khalajhedayati[2], M. Mongillo[1], O. Painter[2], D. Wan[1], A. Potočnik[1], K. De Greve[1,3]

**[1]Imec, Kapeldreef 75, Leuven, B-3001, Belgium**
**[2]AWS Center for Quantum Computing, Pasadena, CA 91106, USA**
**[3]Proximus Chair in Quantum Science and Technology, Department of Electrical Engineering (ESAT), KU Leuven, Leuven, B-3000, Belgium**

## Abstract

Superconducting resonators are central to superconducting quantum information technologies and essential for bosonic qubit architectures, where long-lived storage modes enable hardware-efficient error correction. Achieving ultra-high quality factors in scalable planar circuits is challenging because multiple dissipation channels contribute to the total loss. Here we report planar $\alpha$-Ta resonators fabricated on 300 mm ultra-high-resistivity (>10 k$\Omega$ cm) intrinsic silicon using industrial processes, achieving median internal Q-factors exceeding 40 million and maxima above 60 million. Energy-participation-ratio analysis identify a dominant participation-controlled interface loss mechanism and places conservative upper bounds on substrate-associated dissipation. For the best-performing substrate, the inferred substrate loss tangent is below $1.0 \times 10^{-8}$, establishing industrial MCZ silicon among the lowest-loss substrate platforms reported for superconducting resonators. At the same time, the exceptionally low losses show no clear correlation with commonly cited silicon substrates metrics such as the room temperature resistivity or the impurity concentrations. More broadly, our studies establish industrial 300 mm processing, careful interface engineering and 300 mm MCZ silicon substrates as a promising platform for resonator-heavy superconducting quantum architectures with ultra-high-quality factors.

## Introduction

Superconducting resonators are foundational to modern superconducting quantum technologies: they serve as high-sensitivity probes of microwave loss that limits qubit coherence[1–3] and function as core elements in e.g. cat-qubit–based processors, where bosonic encodings leverage long-lived storage modes for hardware-efficient error suppression.[4–9] To translate these advances into manufacturable quantum systems, there is a premium for planar circuit architectures fabricated on high-resistivity silicon, as they offer straightforward compatibility with semiconductor-grade, high-volume processing and direct pathways to 3D integration.[10–13] In this context, the ongoing quest for ultra-high Q-factors and extended $T_1$ times demands a concerted optimization of materials, fabrication methods, and device designs.[14,15]

Microwave loss in planar superconducting circuits arises predominantly from dielectric/two-level systems (TLS) in the amorphous metal-air, substrate-air, substrate-metal interfaces, as well as losses linked to the bulk substrate.[16–20] Recent work has substantially reduced the surface loss contribution through aggressive wet and dry-cleaning protocols and judicious superconducting metal selection. In particular, α-Ta has emerged as a strong candidate due to both its inherent low oxide loss and its chemical robustness to aggressive etchants that can further reduce surface losses (e.g., hydrofluoric acid, HF).[8,21–24] Such treatments not only reduce $TaO_x$ thickness, but also strip native $SiO_x$ and passivate Si surface bonds, delaying silicon oxide regrowth.[25,26] Follow-up studies have also indicated limitations to such surface treatments; for instance, care must be taken to control hydrogen incorporation in α-Ta films during aggressive etching.[27] By contrast, strategies to suppress substrate–metal interfacial loss, as well as bulk substrate loss, remain comparatively under-explored.[20,28] Their importance has been reinforced by recent studies, where careful selection of substrate and optimization of the interfaces enabled qubit coherence times beyond one millisecond and resonator Q-factors in the 10 million range,[8,29,30] highlighting their critical role in advancing superconducting qubit performance.

Achieving such performance has thus far relied on float-zone (FZ) silicon, which indeed offers crucible-free growth and therefore extremely low background dopant and impurity levels, thereby resulting in ultra-high-resistivity (UHR) substrates (resistivities >10 kΩcm).[31] In contrast, industrial 300 mm fabrication relies exclusively on magnetic Czochralski (MCZ) silicon, which offers large wafer size, mechanical stability, and excellent process uniformity but is understood to inherently incorporate higher concentrations of electrically active impurities, most notably oxygen, carbon, and vacancy-derived defects that limit achievable resistivity.[32,33] Only recently have advances in MCZ crystal growth enabled the fabrication of 300 mm silicon wafers with resistivities in the UHR regime. Whether such impurity-rich MCZ material can nevertheless achieve the ultra-low dielectric loss required at cryogenic temperatures for superconducting quantum circuits remained an open question thus far.

Here we demonstrate α-Ta planar resonators fabricated in a 300 mm industrial process flow on MCZ silicon using a Nb liner (see Figure 1**b,c**) with dilute-HF post-processing, achieving low-power median internal quality factors exceeding 40 million and maximum values above 60 million. Statistical measurements across multiple substrate populations reveal reproducible differences in resonator Q-factors that are neither correlated to nominal wafer resistivity, nor to impurity concentrations detectable with complementary SIMS measurements. Energy-participation-ratio analysis identifies a dominant geometry-dependent dissipation channel, and places conservative upper bounds on substrate-associated loss in industrial MCZ silicon. For the best performing substrate, the inferred upper bound on substrate loss tangent is below $(1.0 \times 10^{-8})$, lower than previously reported estimates for float-zone silicon and sapphire substrates.

# Results

High-Q microwave resonators are fabricated using 300 mm CMOS manufacturing tools in the pilot line facilities of imec (see Methods). Three different 300 mm wafers are used in this study: a high-resistivity (HR) silicon substrate with a resistivity of 3 kΩcm, also used in our previous studies[10,27,34], and two novel UHR wafers with resistivities of 10 kΩcm and 11 kΩcm (Extended Table 1). All three 300 mm diameter wafers are grown using MCZ method in contrast to FZ method typically used for smaller (=< 200 mm) wafers.[19,20] Wafers first undergo diluted HF cleaning to remove native silicon oxide and terminate surface silicon bonds with hydrogen, thereby reducing the rate of subsequent oxide regrowth.[25] After cleaning, they are transferred to the deposition chamber within 8 hours or less (typical transfer time ~1 hour) to minimize oxide growth under ambient conditions. There, a 10 nm Nb film is sputtered to form a liner, followed by deposition of 100 nm Ta at room temperature (RT), all in the same chamber without vacuum break (See supplementary information for additional information on the liner). The RT Ta film grown on 10 nm-thick Nb liner, hereafter referred to as RT Ta for brevity, predominantly grows in the α-phase, exhibiting properties comparable to those reported previously,[35–37] with the Ta film containing less than 1% of β-phase (Supplementary information). After Ta film deposition, 10 nm thick $SiO_x$ is deposited as a sacrificial layer. The wafers are patterned using an industry-standard lithographic process, followed by chlorine-based Ta dry etch as described in our previous work.[34] The patterned designs consist of co-planar waveguide (CPW) resonators and lumped-element (LE) resonators with varying surface participation ratios and resonance frequencies ranging from 4.6 to 6.7 GHz (Figure 1**a** and Supplementary information). Metal etching results in Ta side walls and ~100 nm recess in Si at an angle of ~20° (Figure 1**b**). Energy dispersive spectroscopy (EDS) analysis shows the absence of oxide present at the silicon-niobium and niobium-tantalum interface (Figure 1**c** and Extended Figure 1). After resist stripping and sacrificial layer removal with diluted HF, the wafers are coated with a protective resist layer and diced to a subdie level (samples). After dicing, the protective resist is stripped in acetone and isopropanol, followed by a short oxygen plasma clean to remove organic residues. Samples undergo 1-minute, 10 vol% HF treatment to remove silicon oxide and niobium oxide and reduce tantalum oxide thickness less than 18 hours before the start of a cooldown to millikelvin temperatures. All samples are stored in a nitrogen box, only to be removed during packaging and wire bonding.

Lumped-element resonators fabricated on UHR substrate UHR_10k (Extended Table 1) show very high internal Q-factors ($Q_i$) with a median of 48 million based on measurements collected periodically over 15 hours at single-photon-level power and 10-mK base temperature (Figure 1**d,e**, see Methods for measurement procedure). A power-dependent $Q_i$ is well described by the generalized two-level-system (TLS) loss model[1] (Figure 1**d**). The time-dependent Q-factor data exhibit both global drift as well as distinct jumps, attributed to TLS fluctuations (Figure 1**e**). Resonant frequency fluctuations over time (standard deviation of approximately 56 Hz) show varying degrees of correlation with $Q_i$ fluctuations when sampled simultaneously for the resonator shown in Figure 1**e,f** (Extended Figure 2). Taken together, these observations indicate that the dominant loss mechanism in these very high-*Q* resonators likely originates from TLS-type defects, such as those presumed to be connected to atom tunnelling in disordered materials[1] or trapped quasiparticles.[3,38,39]

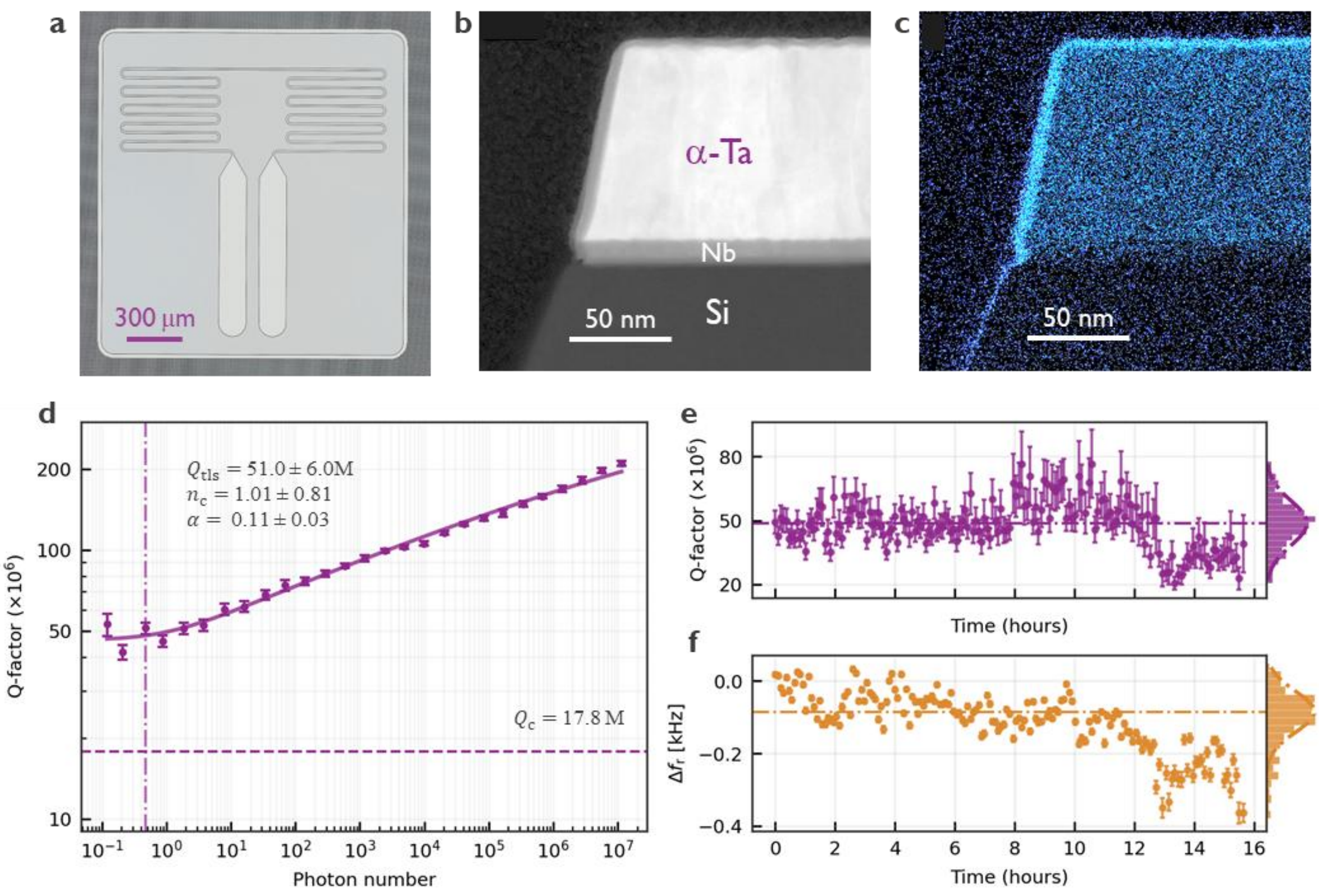

*Figure 1: High Q-factor α-Ta resonators.* ***a*** *Micrograph of a lumped-element resonator with 70 µm gap (LE70) between the coplanar capacitor's pads.* ***b*** *High-angle annular dark-field (HAADF) scanning transmission electron microscope (TEM) cross-section of a patterned sample from the inner edge of a capacitor pad. 100 nm alpha-Ta film is grown on 10 nm thick Nb liner at room temperature. During Ta etch, Si substrate is recessed by ~100 nm.* ***c*** *Energy dispersion spectroscopy (EDS) of the oxygen cross section scan of the same profile as shown in* ***c****. No oxygen can be detected at the Si-Nb-Ta interface.* ***d*** *Power-dependent internal Q-factor of one of the best performing resonators on substrate UHR_10k shows a typical TLS power dependence. Internal Q-factors are plotted against the number of microwave photons in the resonator.* ***e*** *Time dependence of the low power $Q_i$ for the resonator shown in* ***d.*** *Histogram of a time dependent Qi data shows a gaussian distribution with median of Qi = 48 million, typical for TLS.* ***f*** *Time dependence of the resonator's resonance frequency $f_r$ measured simultaneously with the $Q_i$ shown in panel (****e****). Histogram of time dependent $f_r$ is added to the right axis.*

To investigate the role of substrate properties on resonator performance, we compare the TLS loss of resonators fabricated on three different 300 mm MCZ silicon wafers. Specifically, we examined two UHR substrates with resistivities of 10 kΩcm (UHR_10k) and 11 kΩcm (UHR_11k), screened for ultra-high resistivity by commercial suppliers. For reference, we also include a high-resistivity wafer with resistivity in the range of 3-4 kΩcm (REF_3k) as used in our previous studies[25,27,34] (Extended Table 1). To maximize sensitivity to substrate-dependent loss, we focus on LE resonators (LE70, see Supplementary Information) with the lowest energy participation ratio (EPR) (Figure 1**a**, Supplementary information) and subject all samples to diluted HF clean prior to a cool down to minimize surface loss. To improve statistical reliability, the power dependence of $Q_i$ was measured on at least nine resonators from each wafer type. Selected resonators were also chosen for time dependence measurements of at least 15 hours per resonator.

At the single-photon level, the UHR substrates exhibit distinct differences in median $Q_{\mathrm{TLS}}$ (Figure 2**a**). The UHR_10k wafer yields the highest median $Q_{\mathrm{TLS}}$ values of 36.9 million, whereas UHR_11k and REF_3k exhibit comparable median $Q_{\mathrm{TLS}}$ values of 20.1 million and 20.5 million, respectively (Extended Table 1). Time-dependent measurements (Figure 2**b**) further support this trend: UHR_10k exhibits the highest median $Q_{\mathrm{TLS}}$ of 48.9 million, whereas UHR_11k and REF_3k reach best median values of 26.8 million and 27.9 million, respectively. Notably, $Q_{\mathrm{TLS}}$ does not correlate with wafer resistivity for UHR wafers, i.e., the substrate with the highest resistivity (UHR_11k) does not yield the highest median $Q_{\mathrm{TLS}}$, indicating that wafer resistivity alone is not sufficient to predict resonator performance.

Despite a relatively large spread in $Q_{\mathrm{TLS}}$ across devices, statistical analysis reveals a significant difference between substrate groups (Supplementary Information). A one-way ANOVA shows a strong effect of substrate type on $Q_{\mathrm{TLS}}$ ($F = 7.79$, $p = 0.0014$). Post-hoc Tukey comparisons indicate that UHR_10k differs significantly from both REF_3k and UHR_10k, whereas REF_3k and UHR_11k are not statistically distinguishable. The ANOVA effect size ($\eta^2 = 0.275$) indicates that approximately 27% of the total variance in $Q_{\mathrm{TLS}}$ across devices can be attributed to differences between substrate types, with the remaining variance arising from device-to-device variation within a given wafer. The existence of a statistically significant substrate-dependent effect suggests that substrate loss differs between the three wafers. To test this hypothesis, we examine the scaling of TLS loss with surface participation ratio across the REF_3k and UHR_10k substrates, as these represent statistically distinct groups.

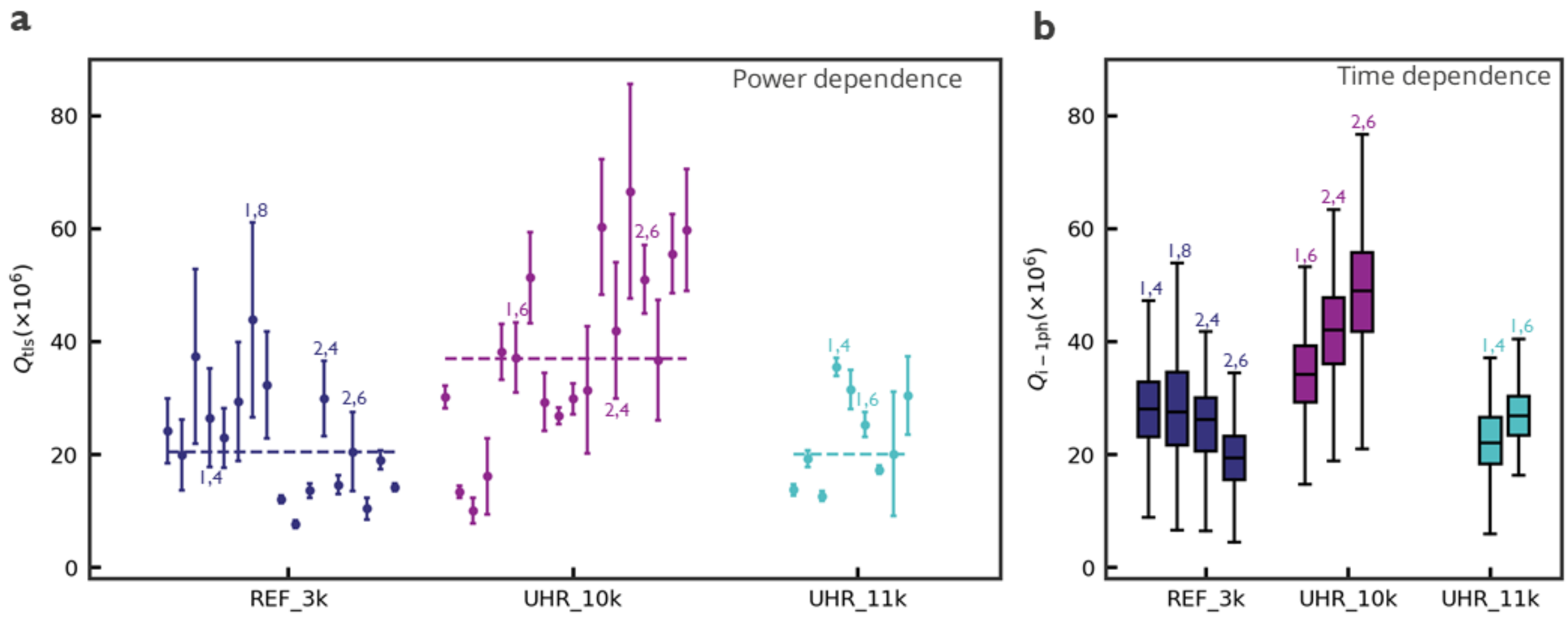

*Figure 2: Q-factors as a function of Si substrate type.* ***a*** *$Q_{TLS}$ extracted from the power dependence of individual resonators fabricated on a high-resistivity silicon substrate (REF_3k) and two ultra-high-resistivity silicon substrates (UHR_10k and UHR_11k). Points show values obtained from fits to the power dependence of each resonator; error bars indicate fitting uncertainties. Labels above the points denote the sample and resonator indices (sample number, resonator number). Horizontal dashed lines denote the median of each substrate group.* ***b*** *Time dependent low power $Q_i$ measured at ~1 photon for representative resonators from each substrate. Boxes indicate the interquartile range, whiskers the full data range, and centre lines the median. UHR_10k shows the highest median $Q_{TLS}$ (48.9 million), while UHR_11k and REF_3k yield similar best median values of 26.8 million and 27.9 million, respectively.*

The TLS loss in the resonator consists of a surface component that scales with resonator geometry and a bulk component that is largely geometry independent. This relationship can be expressed as

$$\frac{1}{Q_{\mathrm{TLS}}} = p_{\mathrm{surface}} \tan\delta_{\mathrm{surface}} + p_{\mathrm{Si}} \tan\delta_{\mathrm{Si}}, \qquad (1)$$

where surface loss combines metal-substrate, metal-air and substrate-air interface losses, represented by the product of an effective surface participation ratio $p_{\mathrm{surface}}$ and an effective surface loss tangent $\tan\delta_{\mathrm{surface}}$[10,22,29]. The geometry-dependent surface contribution is commonly represented using the metal–substrate participation ratio as an effective parameterization of the combined surface-interface losses[10,22,29]. Similarly, bulk substrate loss is given by the product of substrate participation ratio $p_{\mathrm{Si}}$ and the substrate loss tangent $\tan\delta_{\mathrm{Si}}$. Within this decomposition, $\tan\delta_{Si}$ acts as an effective parameter for any contribution that remains independent of the chosen surface participation ratio.

Two complementary statistical analyses are used to examine the measured TLS loss ensemble (Figure 3). We first analyse the median TLS loss for each resonator geometry, which characterizes the typical device realization while preserving information contained in the broader ensemble (Figure 3**a**). Both substrate populations exhibit similar participation-dependent scaling, consistent with a common dominant surface-loss mechanism. However, the REF_3k and UHR_10k datasets remain systematically separated at low participation ratios, where the geometry-dependent contribution is reduced. Within the framework of Eq. (1), this separation corresponds to effective substrate loss tangents of (2.31 ± 0.76) x$10^{-8}$ for REF_3k and (0.64 ± 0.33) x$10^{-8}$ for UHR_10k.

To assess the robustness of this participation-independent separation, we additionally analyse the lower-loss envelope of the resonator ensemble using the minimum measured TLS loss within each substrate population (Figure 3**b**). This approach probes the intrinsic dissipation floor under the assumption that additional process-dependent loss mechanisms can increase the measured loss of individual devices without reducing the underlying participation-controlled limit. In this limit, no systematic separation is observed between the two substrate populations, and the combined dataset is well described by the surface-loss term of Eq. (1) alone, without the need for the geometry-independent substrate component.

Despite the different treatment of the participation-independent contribution, both analyses recover nearly identical participation-dependent scaling. The lower-envelope analysis yields an effective $\tan\delta_{\mathrm{surface}}$ of $(1.81 \pm 0.07) \times 10^{-4}$, while the median analysis yield $(2.02 \pm 0.25) \times 10^{-4}$ for REF_3k and $(2.02 \pm 0.14) \times 10^{-4}$ for UHR_10k. The close agreement between these values indicates that the dominant participation-dependent dissipation is largely independent of the statistical treatment of the dataset. Independent participation-ratio anchoring analysis indicates that the metal-substrate interface is more likely the dominant surface-loss channel (Supplementary Information). Assuming all surface loss comes from the metal-substrate interface, the above surface loss tangent values provide a reliable upper bound for $\tan\delta_{\mathrm{MS}}$ of $\sim 2.3 \times 10^{-4}$, which is comparable to the lowest reported values[22,24,29,37].

The differing behaviour of the median and lower-envelope statistics indicates that the effective $\tan \delta_{Si}$ term is not uniquely determined by the present measurements. One possibility is that it reflects a substrate-associated dissipation channel that becomes observable once the dominant participation-dependent scaling is removed. Alternatively, it may arise from process-dependent or statistical variations that are absorbed into the effective substrate term without affecting the participation-controlled limit. Both interpretations remain consistent with the present dataset. Consequently, the median and lower-envelope analyses are best viewed as providing complementary bounds on the admissible loss landscape. Under this interpretation, attributing the entire effective $\tan \delta_{Si}$ term to substrate dissipation yields conservative upper bounds on the effective substrate loss tangents of 3.1 x$10^{-8}$ and 1.0 x$10^{-8}$ for REF_3k and UHR_10k, respectively.

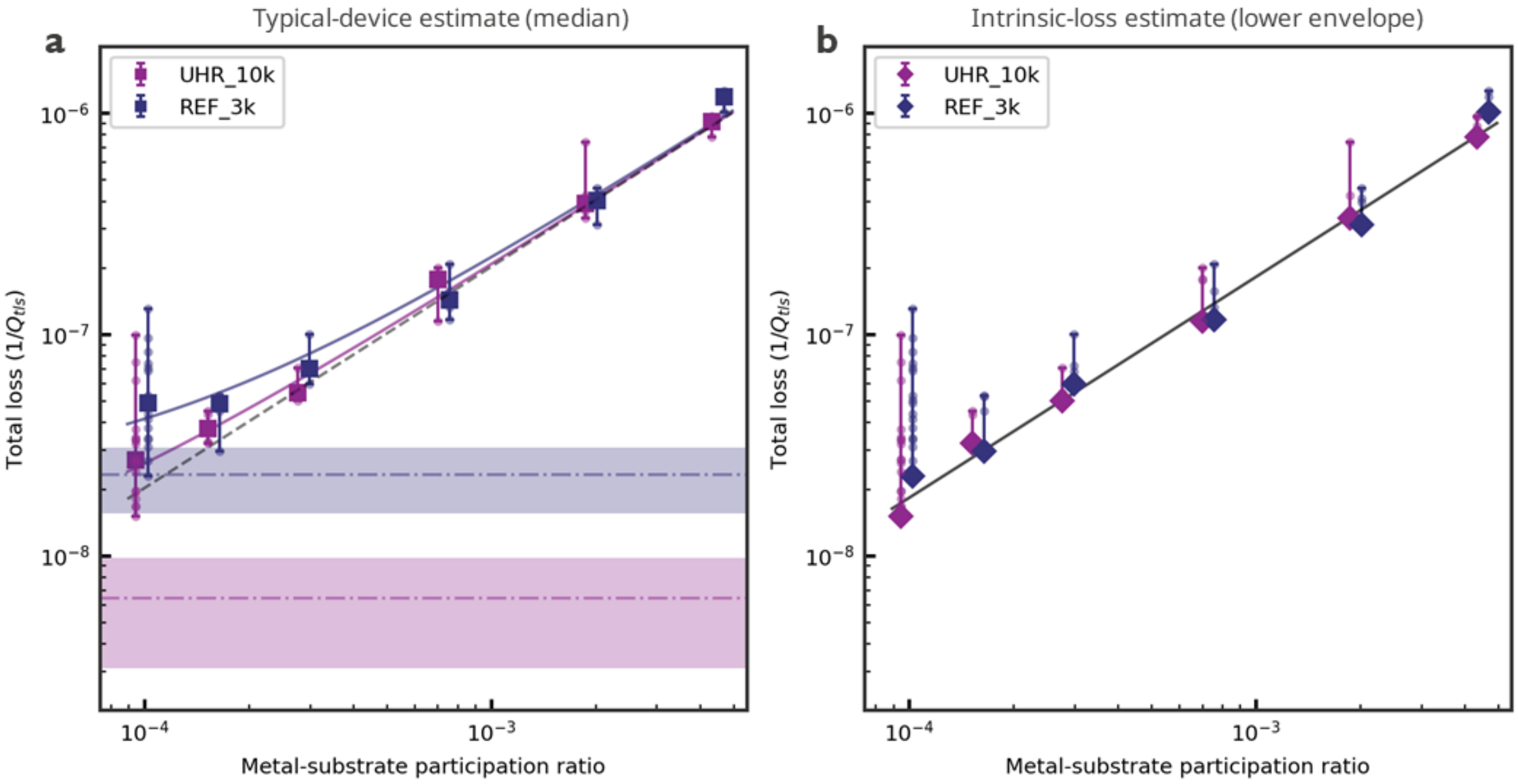

*Figure 3: Energy participation ratio scaling of the microwave loss as a function of metal-substrate participation ratio for resonators fabricated on REF_3k and UHR_10k substrates. Individual resonator measurements are shown as circles. Large symbols denote the median (a) or minimum (b) loss extracted for each geometry. Solid lines are fits performed using the appropriate form of Eq. (1) as described in the text.* ***a****. shows the typical resonator realization and preserves ensemble statistics.* ***b****. lower-loss tail of the resonator distribution provides an estimate of the intrinsic participation-controlled dissipation floor. Both analyses recover similar participation-controlled scaling, while the median data exhibit a systematic substrate-dependent offset that may indicate an additional geometry-independent loss contribution. Error bars correspond to the fit parameter uncertainty.*

To investigate whether residual impurities in the substrate account for the substrate-dependent behaviour observed in the resonator ensemble, we perform dynamic SIMS measurements of boron, carbon, nitrogen, phosphorus and oxygen on the MCZ silicon used in this study (Figure 4). For comparison, we additionally analysed a FZ silicon wafer (>20k kΩcm) used in Ref. [37], where high-Q tantalum resonators have previously been demonstrated. From the measured elemental concentrations, we can observe that the FZ wafer exhibits generally lower impurity concentrations than the MCZ substrates, most notably for oxygen, which is nearly two orders of magnitude higher in the MCZ material and B which is below the detection limit for the FZ wafer, both consistent with expectations for the different growth techniques. Less pronounced changes can be detected for within MCZ wafers, while nitrogen and phosphorus concentrations are below the detection limit of the measurement for all wafers. Despite these observed differences in impurity content and growth method, we find no systematic dependence of the extracted resonator quality factors on the concentration of any measured species within the sensitivity of the measurement.

The above observations demonstrate that neither nominal resistivity nor conventional impurity metrics captured by SIMS are sufficient to explain the observed microwave loss. Within the range of dopant concentrations investigated here, which remain near or below the SIMS detection limit, the substrate-dependent behaviour inferred from the ensemble statistics in Figure 3 does not correlate with any measured impurity concentration. If a substrate-associated loss channel is indeed present, its origin therefore appears to be governed by a parameter not captured by standard resistivity specifications or elemental impurity concentrations. More broadly, the results indicate that microwave dissipation cannot be predicted from conventional measures of substrate quality alone. While the ensemble statistics reveal a substrate-dependent signature, the present measurements do not uniquely identify its microscopic origin, and neither nominal resistivity nor the impurity concentrations measured by SIMS provide a predictive metric for resonator performance. The dominant participation-controlled dissipation and any additional substrate-correlated contribution therefore appear to be governed by material or process-dependent factors that are not captured by standard substrate characterization. These conclusions should not, however, be extrapolated to substrates with substantially higher electrically active dopant concentrations, where carrier-mediated microwave absorption may become significant.

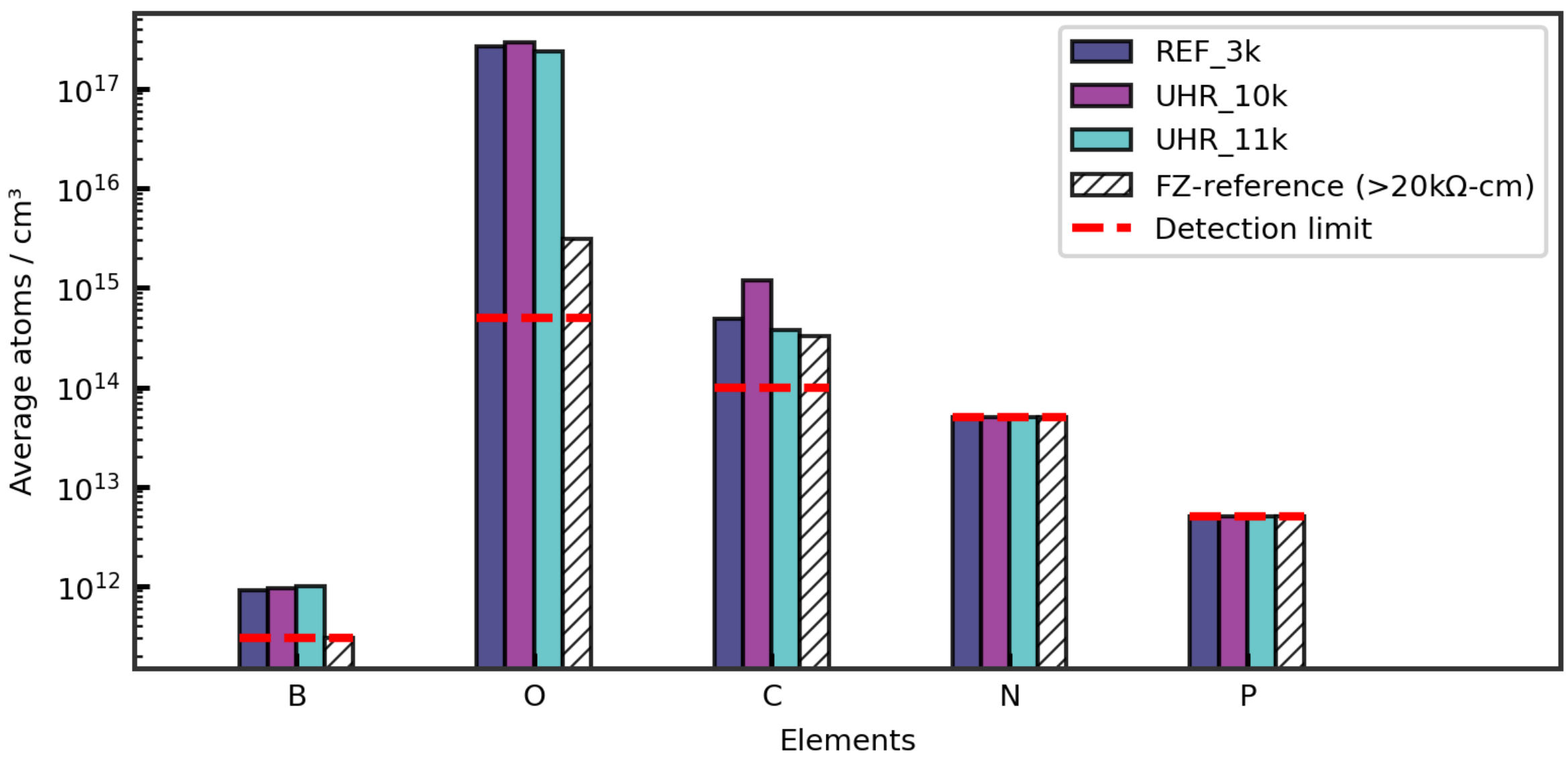

*Figure 4: Impurity concentrations in silicon substrates measured by dynamic SIMS. Average concentrations of B, O, C, N and P measured by dynamic SIMS for the MCZ wafers used in this study (REF_3k, UHR_10k and UHR_11k) and a float-zone (FZ) reference wafer (Ref. 37). The FZ substrate shows generally lower impurity concentrations, particularly for oxygen. Dashed lines indicate the detection limit; N and P are below this limit for all wafers.*

## Discussion and conclusions

Our results establish industrial 300 mm MCZ silicon as a viable platform for ultra-high-coherence superconducting resonators. Upper bounds on the substrate loss tangent of 1.0 $\times 10^{-8}$ for UHR_10k and 3.1 $\times 10^{-8}$ for REF_3k are comparable to the lowest reported estimates for HEMEX sapphire substrate [$(1.9 \pm 0.6) \times 10^{-8}$][8], and ultra-high-resistivity wafer (20 kΩcm) grown FZ Si [$(1.9 \times 10^{-8})$][29], highlighting the exceptional microwave performance achievable with industrial MCZ substrates (Extended Figure 3).

Despite substantial differences in impurity concentrations between substrates and crystal-growth methods, dynamic SIMS measurements reveal no systematic correlation between resonator loss and any measured impurity species. Combined with the absence of a clear dependence on nominal wafer resistivity, these observations indicate that microwave dissipation in high-purity silicon cannot be predicted from conventional substrate metrics alone and is governed by additional material or process-dependent factors.

From a device-engineering perspective, the most important implication of this work is that substantial headroom for further improvement remains at the interfaces. Importantly, this conclusion is robust to the statistical treatment of the resonator ensemble: both lower-envelope and median-based analyses recover nearly identical participation-dependent scaling and attribute a substantial fraction of the measured loss to interface-related dissipation. Consequently, irrespective of the microscopic origin of the residual component, reducing losses associated with the remaining $TaO_x$ and Si-Nb-Ta interfaces represents the most direct route towards higher resonator coherence.

Beyond material and process optimizations, the demonstrated high-Q resonators are immediately applicable to integrated superconducting quantum processors, for example those employing bosonic-code quantum error correction.[7] Most importantly, these exceptionally high Q-factors were achieved through advanced, high-volume fabrication methodologies in an industry standard 300 mm CMOS pilot line, ensuring reproducibility and scalability. The planar processes employed are inherently compatible with 3D integration strategies essential for realizing large-scale resonator-heavy quantum computing architectures[40].

## Methods

High-Q microwave resonators are fabricated in the 300 mm CMOS pilot line facilities of IMEC. State-of-the-art industrial fabrication tools are used for the resonator fabrication process. This includes stepper based DUV 248 nm-lithography, an Inductively Coupled Plasma (ICP) etch system where etching is performed using $Cl_2/CH_2F_2$ gasses at 13.56 MHz RF frequency, and industrial cleaning tools and wafer dicing – all state of the art, shared tools, commonly used for 300 mm CMOS pilot line fabrication.

Industry-grade high- and ultra-high-resistivity 300mm diameter wafers grown by the magnetic Czochralski method (MCZ)[32] contain intrinsic boron p-type impurities, have a thickness of 775 μm, exhibit fewer than one surface light point defect (LPD) per wafer, and show oxygen concentrations ranging from approximately 3 ppma (UHR_11k) to about 10 ppma (REF_3k,

UHR_10k) (Extended Table 1). Despite the somewhat higher oxygen content in REF_3k and UHR_10k wafers, thermal oxygen donors are unlikely to form during resonator fabrication because the processing temperature remains low (<200 °C).[41]

Wafer specifications provided by vendors (Extended Table 1) are determined from measurements on two slugs (thick cylindrical samples) cut from the beginning and end of the region of interest in the silicon ingot. Electrical resistance is measured using four-probe technique, impurity concentrations are determined by Fourier-transform infrared spectroscopy (FTIR), and light particle density (LPD) is assessed using dark-field optical scattering method.

Fabricated, diced, and cleaned samples (subdies) are packaged and wire-bonded directly to a custom-made aluminium (AW-1350A) sample holder. The sample ground plane is wire-bonded to the package, and the feedline is bonded directly to a non-magnetic SMP bulkhead connector (Supplementary information). During bonding, the sample is kept in place by vacuum; after bonding, it is secured by wire bonds themselves. No adhesive is used in the package. Additional on-sample wire bonds are added to connect ground planes and minimize package modes.

Between the HF post-processing step and cooldown, the samples are stored in the nitrogen storage box with $N_2$ gas flow of 10 scch and a gas purity grade of N8.0.

Wire-bonded packaged samples are mounted on to a vertical copper bracket attached to the mixing plate of a dilution refrigerator and enclosed in a cryo-perm magnetic cylindrical shield with an open top. The packages are positioned more than 7 cm below the shield's open end. A second magnetic shield is installed at room-temperature in the vacuum can of the dilution refrigerator.

Resonators are characterized using a vector network analyser (VNA, Keysight P5004). The microwave power reaching the samples at the mixing chamber plate, attenuated by ~87 dB, was previously calibrated using a Transmon qubit. This calibrated value is used to estimate the microwave photon number inside the resonator. In addition to the setup described in our previous work, we include a variable attenuator at the room-temperature stage, set to 20 dB, with an insertion loss of approximately 7 dB. The overall experimental setup and measurement procedure otherwise follow those described in our previous publications.[34,27] We note that the VNA integration bandwidth (IF_BW) must be set below 10 Hz to reliably measure the resonance linewidth at high photon numbers ($>10^8$) where Q-factors exceed 100 million.

Q-factor measurements are performed on lumped-element (LE) resonator with the capacitor electrode gap of 70 μm (electrode width of 150 μm) and coplanar-waveguide (CPW) resonators with central trace width (=gap) of 1 μm, 3 μm, 10 μm, 30 μm, and 60 μm. We refer to the supplementary information for more information on their design and simulated energy participation ratios.

# Acknowledgments

The authors gratefully acknowledge Olivier Richard, Patrick Carolan, Laura Nelissen and Alex Merkulov for metrology support, and Erwin Vandenplas and Koen Verhemeldonck for wirebonding support. This work was supported in part by the imec Industrial Affiliation Program on Quantum Computing. The work was also performed under JDP with imec, Kapeldreef 75, 3001 Leuven, Belgium. The authors thank Guillaume Marcaud, Matthew Hunt and Rassul Karabalin for insightful comments on this work. Authors also thank Simon Lapointe for helping with 2D simulations in Palace.

# Contributions

R.A., D.P.L. and A.P. conceived and designed the experiment. T.I., S.M., C.V., Y.C., and Y.L. contributed to the device fabrication. R.A., and A.M.V., participated in the measurements. D.P.L and A.H. performed SIMS measurements. R.A. and S.A. performed microwave simulations. R.A., D.P.L, J.V.D, O.P and A.P analysed the data. R.A, D.P.L. and A.P. led the paper writing, and all other authors contributed to the text. A.P, M.M, D.W and K.D.G supervised the project.

# Extended Material

**Extended Figure 1**: Scanning Tunnelling Electron Microscopy (STEM) and Energy Dispersion Spectroscopy (EDS) images of α-Ta resonator cross section from the inner edge of the lumped-element capacitor shown in Figure 1**a**. **a** High-Angle-Annular-Dark-Field (HAADF)-STEM cross section of Ta resonator. The coloured square marks the area where high magnification image is taken. **b** High magnification DF-STEM cross section of the M-A interface. Lines indicate Nb liner thickness and Nb-Si intermixing thickness. **c** Tantalum, **d** nitrogen, **e** niobium, **f** fluorine, **g** oxygen and **h** chlorine EDS maps of the cross-section shown in **a**. No N, F, or Cl impurities can be detected at the interfaces. As previously reported[34], the difference in $TaO_x$ thickness between the sidewalls and the top surface results from the different effective oxygen plasma exposure experienced by these regions during resonator fabrication and cleaning processes.

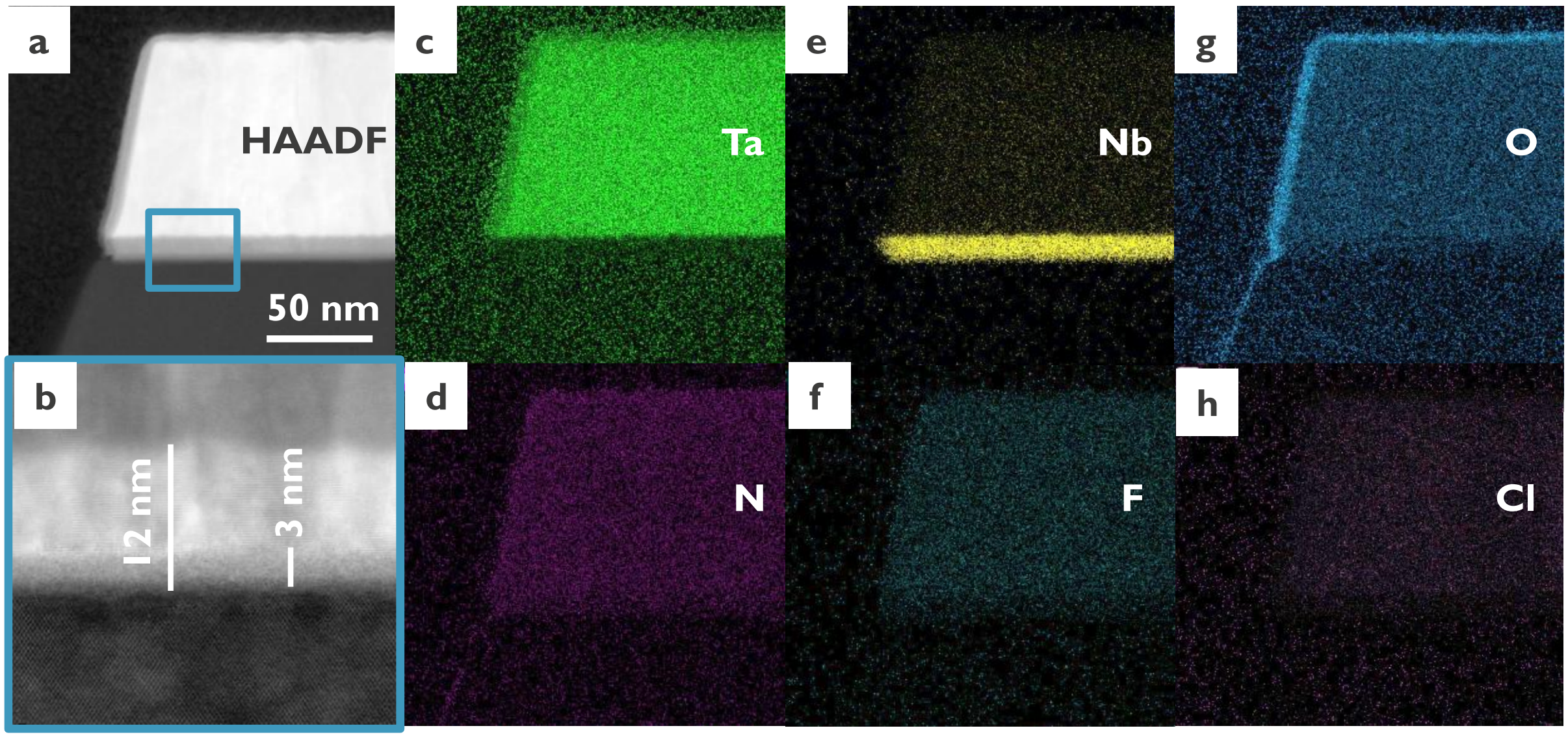

**Extended Figure 2**: Correlation between resonator $Q_i$ and resonator frequency fluctuations. **a** Time evolution of the internal quality factor $Q_{\mathrm{i-lp}}$(purple) and the resonator frequency fluctuation $\Delta f_{\mathrm{r}}$(orange) measured over $\sim 16$ hours for the resonator shown in Figure 1**d,e,f**. Error bars denote the uncertainty obtained from the resonator fits. **b** Time-resolved Pearson correlation coefficient $r$ between $Q_{\mathrm{i-lp}}$ and $\Delta f_{\mathrm{r}}$, calculated using a sliding time window along the data shown in **a**. The black curve shows the Pearson correlation coefficient $r$ as a function of time, while the grey curve indicates the corresponding $p$-value of the correlation test (right axis). The shaded region marks the initial time interval required to accumulate sufficient data points for the first correlation estimate, determined by the size of the sliding window used in the analysis. While the correlation fluctuates at earlier times, a pronounced positive correlation ($r \gtrsim 0.6$) emerges at later times ($\gtrsim$ 12 h), indicating that fluctuations in $Q_{\mathrm{i-lp}}$ and $\Delta f_{\mathrm{r}}$ become strongly coupled during this period.

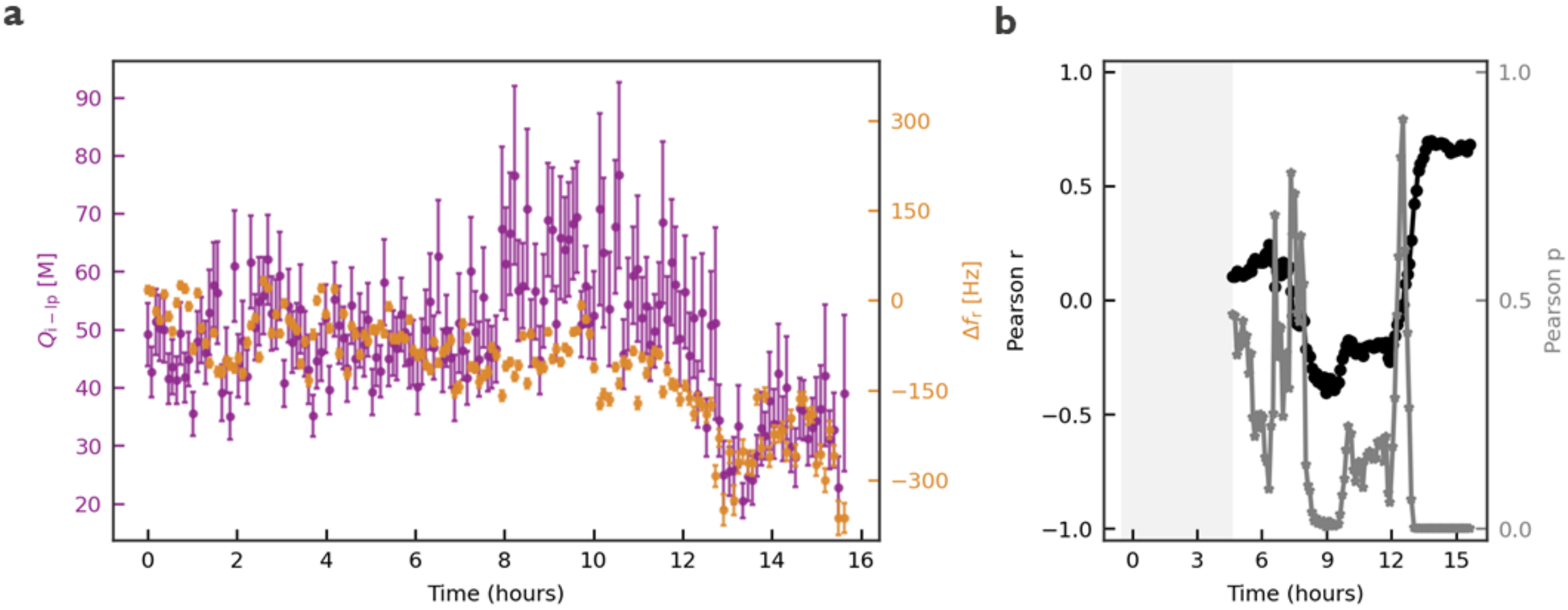

**Extended Figure 3**: Comparison of substrate loss tangents reported in literature. [17,18,42] use a combination of resonator measurements EPR scaling and Monte-Carlo simulations, [20] uses an extrapolation based on loss scaling with bulk doping concentrations (determined at the undoped measurement value), [16,29] uses qubit measurements with EPR scaling and the resulting saturation, [28] a high-quality microwave 3D cavity mode measurements with in-situ insertion of dielectric substrates, [8] uses multi-mode resonator measurements to disentangle substrate and surface loss. Downward arrows indicate conservative upper bounds for this work, whereas literature values correspond to reported extracted substrate loss tangents.

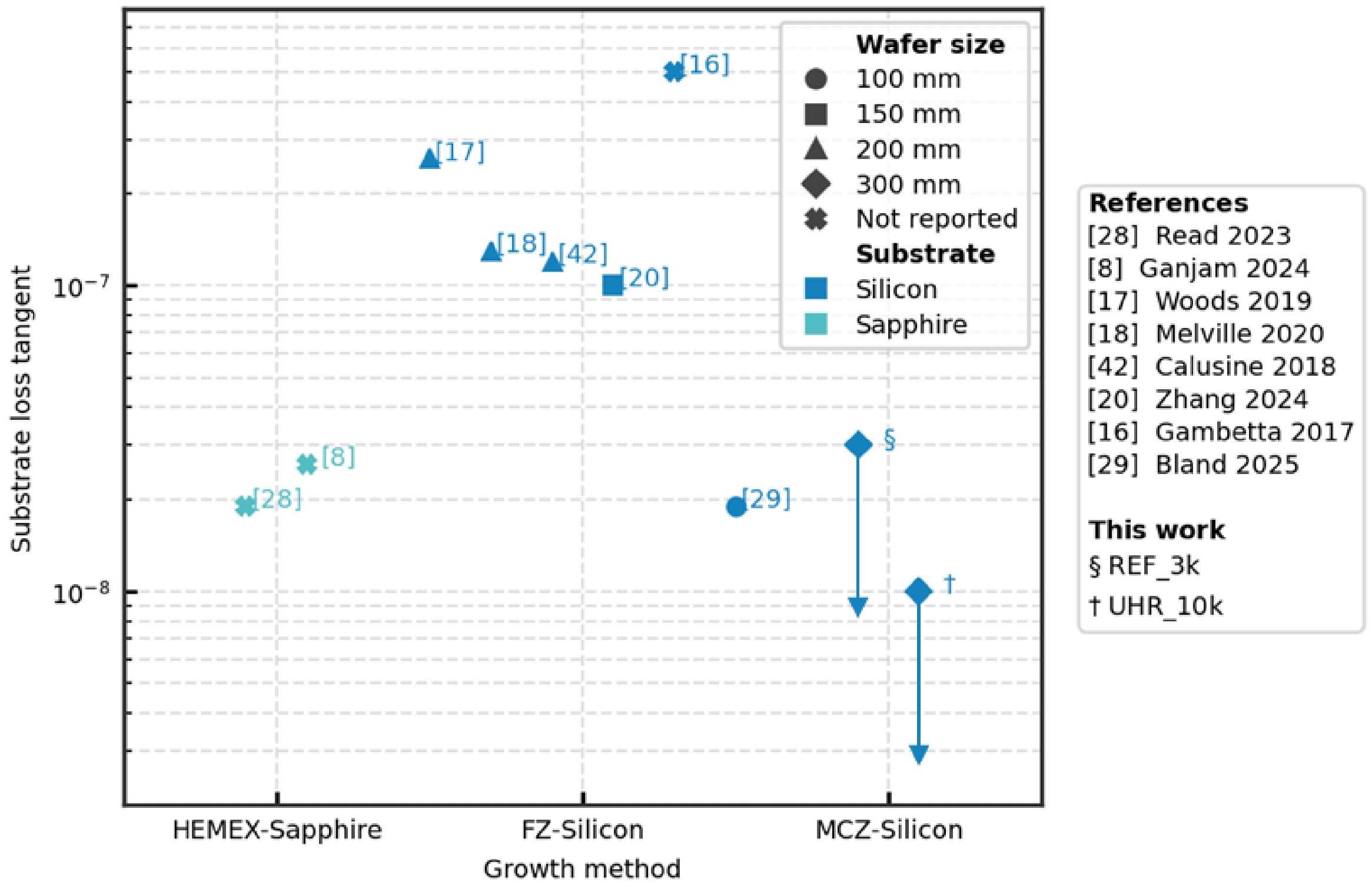

**Extended Table 1**: Specification comparison of the four tested Si substrates. Substrate electrical resistivity measured at RT, oxygen content and light point defects (LPD) count are provided by substrate vendors, resonator internal Q-factors are measured in this work as shown in Figure 2**a**. All substrates share the following properties: intrinsic p-type doping (B), diameter: 300 mm, thickness: 775 μm, double-side polished, <100> crystal orientation, growth method: Magnetic-Czochralski. Resistivities indicated in the labels and in the main text correspond to the lowest rounded measured resistivities.

| **Label** | **Resistivity (kΩ cm)** | **Oxygen (ppma)** | **Carbon (ppma)** | **Mean LPD >90 nm (ct/wfr)** | **$Q_{TLS}$ (median/max)** |
|---|---|---|---|---|---|
| REF_3k | 3.3-4.1 | 8-10 | <0.5 | 0-0.2 | 20.5 M / 43.8 M |
| UHR_10k | 9.6-9.8 | 8-10 | <0.5 | 0.56 | 36.9 M / 66.6 M |
| UHR_11k | 11-13 | 3.1-3.3 | 0.001 | 0.36 | 20.1 M / 35.5 M |

**Extended Table 2**: Substrate loss estimates. Fit results from Figure 3**a** using $p_{\mathrm{Si}} \sim 0.92$.

| **Wafer label** | **Effective surface loss tangent ($\tan\delta_{\mathrm{surface}}$)** | **Substrate loss tangent ($\tan\delta_{Si}$)** | **Substrate loss fraction** |
|---|---|---|---|
| REF_3k | $(2.02 \pm 0.25) \times 10^{-4}$ | $(2.31 \pm 0.76) \times 10^{-8}$ | $0.54 \pm 0.18$ |
| UHR_10k | $(2.02 \pm 0.14) \times 10^{-4}$ | $(0.64 \pm 0.33) \times 10^{-8}$ | $0.25 \pm 0.13$ |

# Supplementary Information

## α-Ta film characterization

The Ta film (100 nm) with a 10 nm Nb liner, deposited on HF-cleaned Si substrates at room temperature, is characterized by X-ray diffraction (XRD) measurements at multiple locations across the wafer using a θ–2θ geometry. The XRD pattern shows a dominant peak at 38.51° (Figure S1), corresponding to α-Ta (110), confirming the desired Ta phase across the wafer. A much weaker peak appears at 34.7°, roughly five orders of magnitude lower in intensity, which we attribute to β-Ta (002). Previous studies have reported that Ta films grown at room temperature with Nb seed layers on Si and sapphire can contain a small fraction of β-Ta.[43] Although the expected position for β-Ta (002) is 33.77°,[36] compressive strain in the film likely shifts this peak toward higher angles. We also rule Nb as the origin of this peak, since the closest Nb reflection - Nb (110) at 38.37°- would overlap with the α-Ta peak. Following the methodology used in Ref. [44], we estimate the β-Ta volume fraction to be approximately 0.8%.

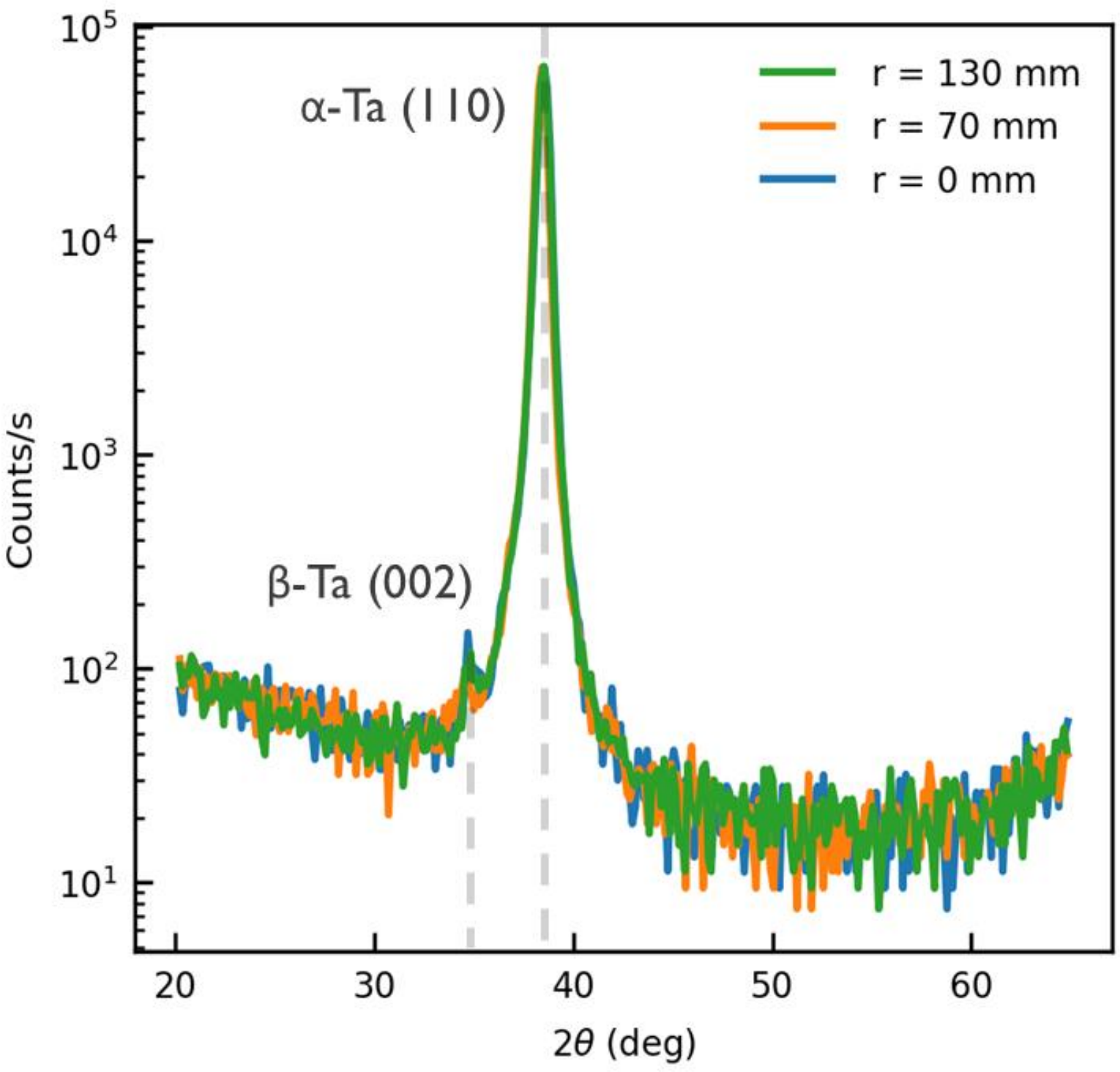

*Figure S1: XRD spectra measured at three different locations, r = 0 mm (center), r = 70 mm (middle), r = 130 mm (edge) from the center of the wafer.*

To independently verify the phase of RT Ta we next investigate Ta film resistivity. First sheet resistance wafer maps are measured with a KLA Tencor RS100 four-point system (Figure S2). A linear array of probes with equal spacing was used for the measurements, and two configurations were used in the measurements as shown in Figure S3. Using the values obtained sheet resistance ($R_s$) is calculated using the following formula:

$$R_s = \frac{\pi}{\ln 2} \frac{R_a^2}{R_b}$$

where $R_a$ is the resistance measured using configuration A and $R_b$ is the resistance measured in configuration B.

The film resistivity is then calculated using the sheet resistance and the nominal deposited film thickness (10 nm Nb + 100 nm Ta). The mean resistivity across the entire wafer is 20.45 ± 0.06 μΩcm, which is consistent with the resistivity of bulk α-Ta.[34] Less than 1% of β-Ta will not affect extracted film resistivity.

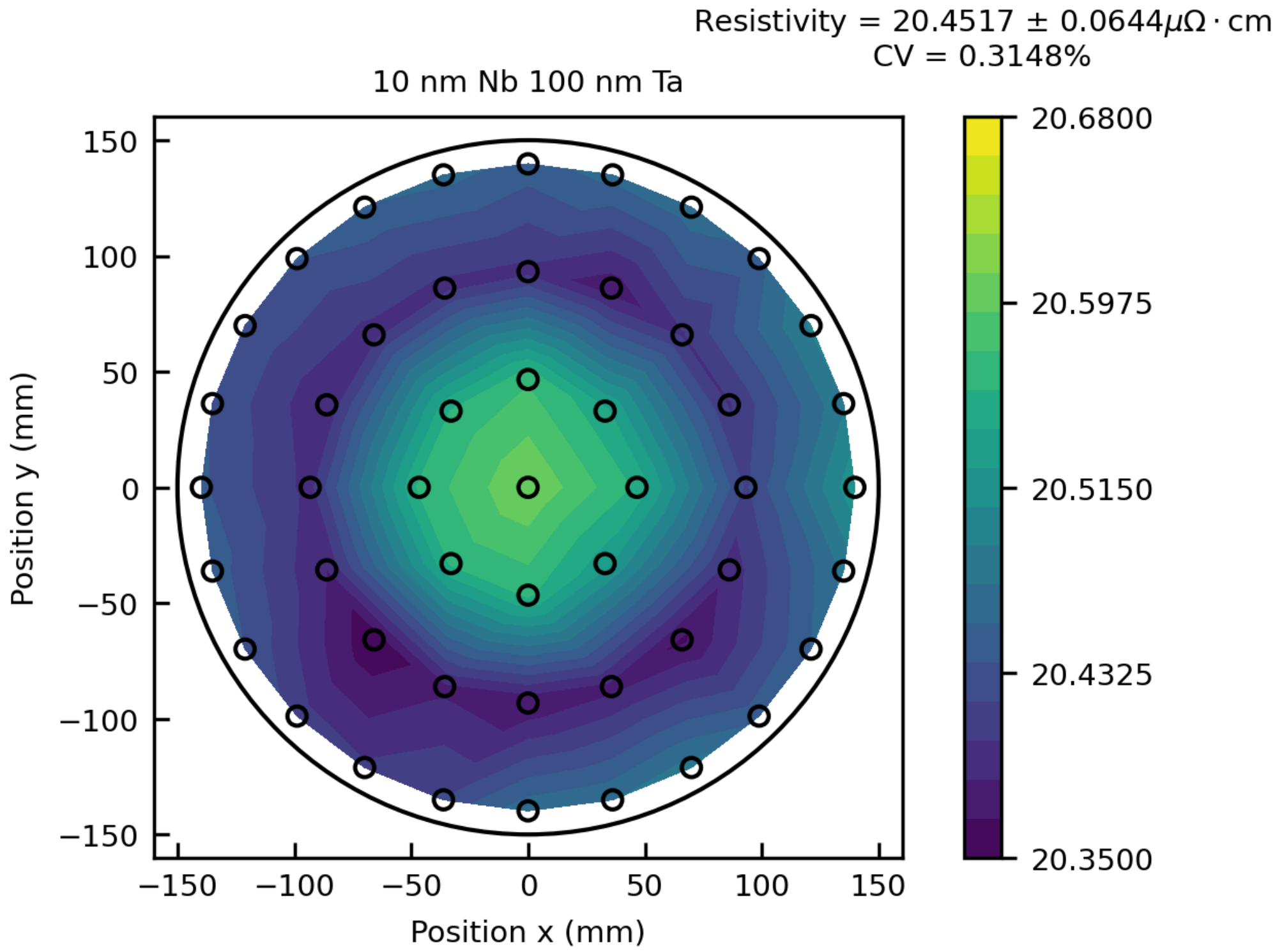

*Figure S2. Resistivity map of the Ta-Nb film used for high-Q resonator fabrication.*

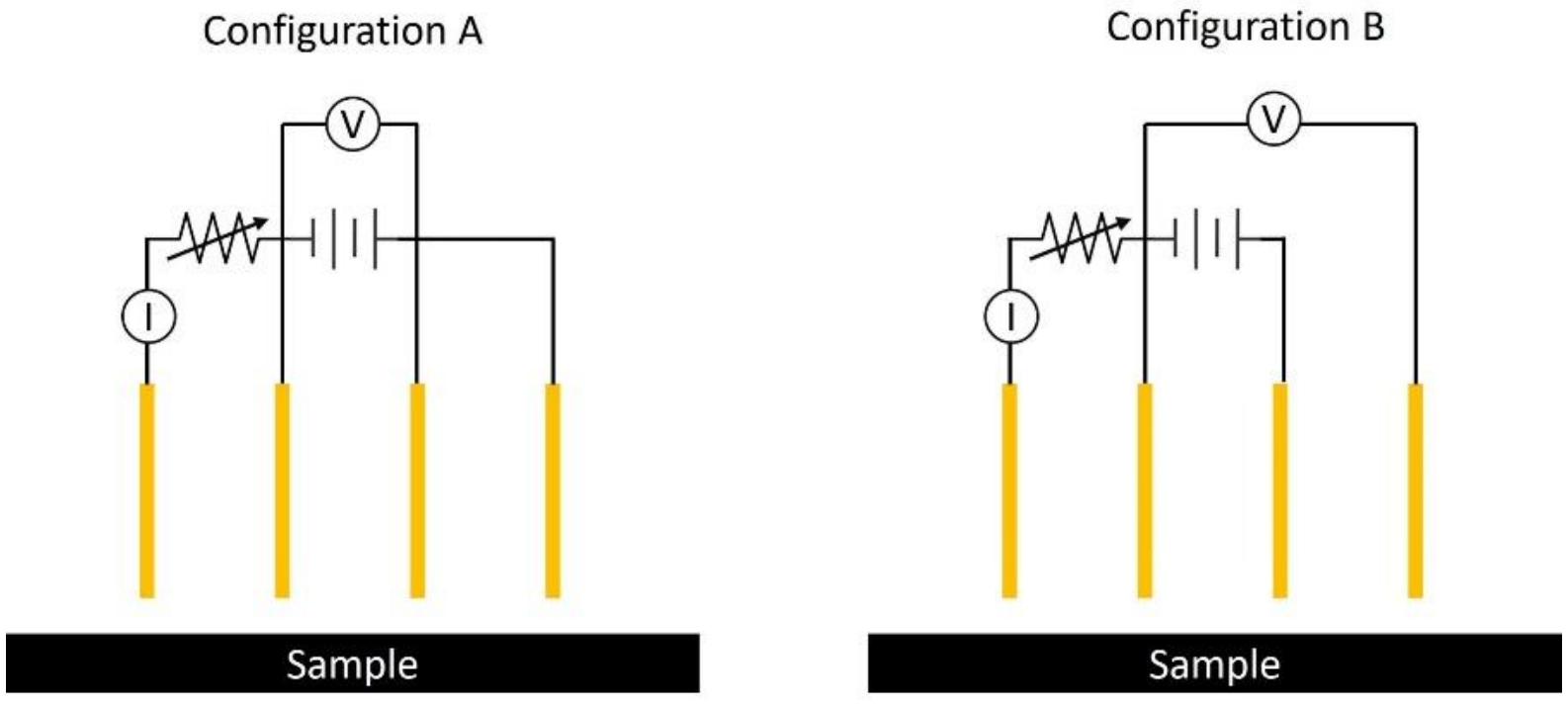

*Figure S3. The two configurations used to measure wafer-scale sheet resistance at room temperature.*

AFM measurements (Figure S4) were performed to examine the surface topology of the RT Ta film. Measurements were taken at the wafer center ($r$ = 0 mm) and at the wafer edge ($r$ = 120 mm). The $R_q$ (root mean square roughness) is 0.533 ± 0.04 nm, indicating that even ~100 nm Ta films can achieve smooth surfaces. The mean $S_z$ (vertical range between highest and lowest point measured) is 5.25 nm, suggesting low topographical variation. Additionally, the surface morphology exhibits distinct worm-like features, consistent with previous reports for similar Nb-Ta layers.[37]

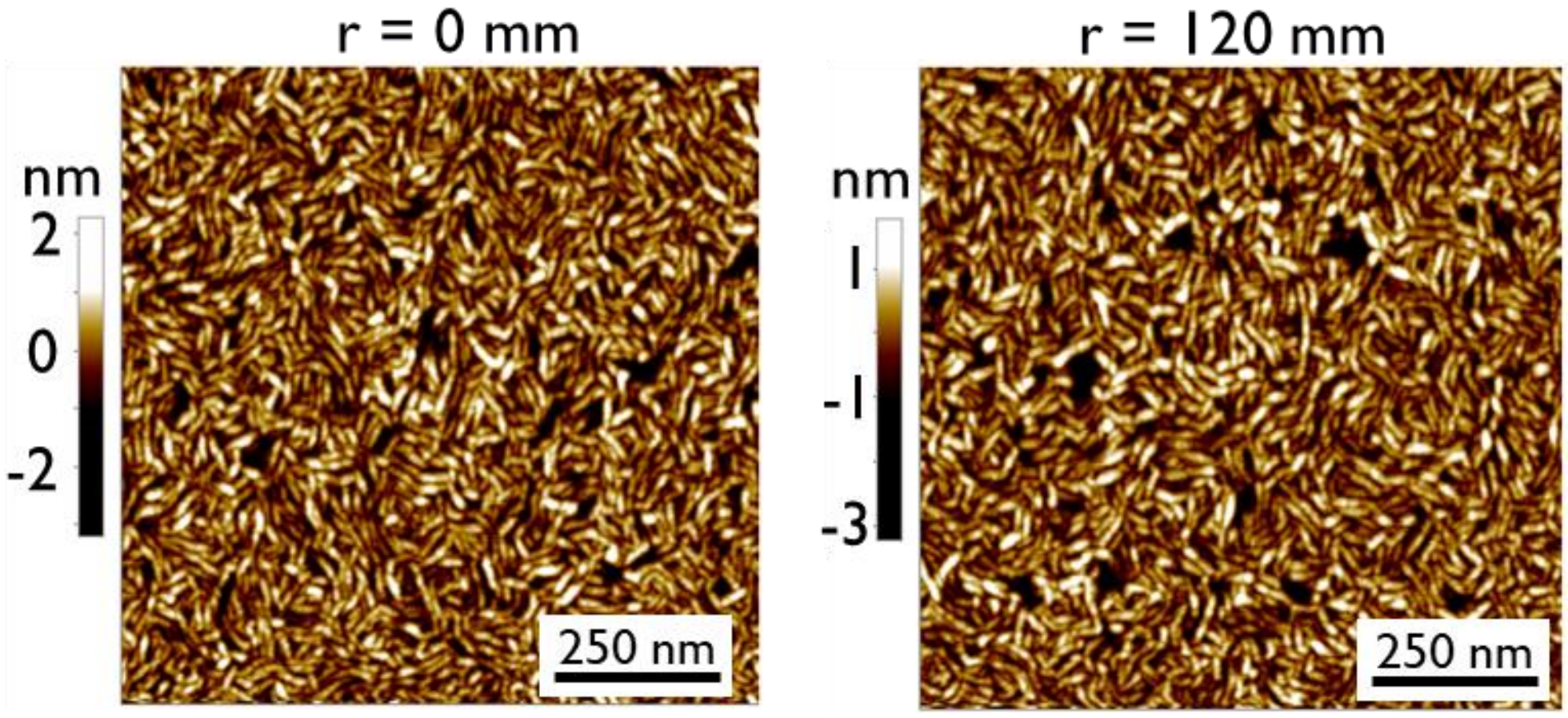

*Figure S4. AFM surface topology of RT Ta films used for high-Q resonator fabrication.*

## Resonator designs

The device set comprises of lumped-element (LE) resonators (Figure S5**a,b**) with capacitor gap of 70 µm (LE70. Each capacitor pad measures 150 µm in width and 800 µm in length, while the inductive meander has a line width of 10 µm. Meander lengths are varied to achieve resonant frequencies between 4.6 GHz and 6.7 GHz. Each LE sample contains nine resonators whose meanders are inductively coupled to a common feedline. Experimentally, we obtain $Q_c$ values between 1-25 million.

We implement coplanar waveguide (CPW) λ/4 resonators (Figure S5**c,d**) with geometries W1, W3, W10, W30, W60 and W120, where 'W' denotes both the center trace width in µm and the gap to the ground plane. Each sample comprises 12 CPW resonators, with 2 resonators of a given geometry, capacitively coupled to a common feedline. The lengths of the CPWs are varied to achieve resonance frequencies between 4.6-6.6 GHz. Experimentally measured coupling quality factors ($Q_c$) fall within the ranges 1–2 million for W1, 1–2 million for W3, 2-4 million for W10, 7-40 million for W30, 7-40 million for W60 and 1-2 million for W120. For the W120 designs, the experimentally obtained $Q_i$ exceeds $Q_c$ by more than an order of magnitude. Therefore, we exclude these data points in Figure 3, although their inclusion does not affect the conclusions of the main text. The feedline of both LE and CPW samples has a center trace width of 10 µm and the gap to the ground plane is 6 µm.

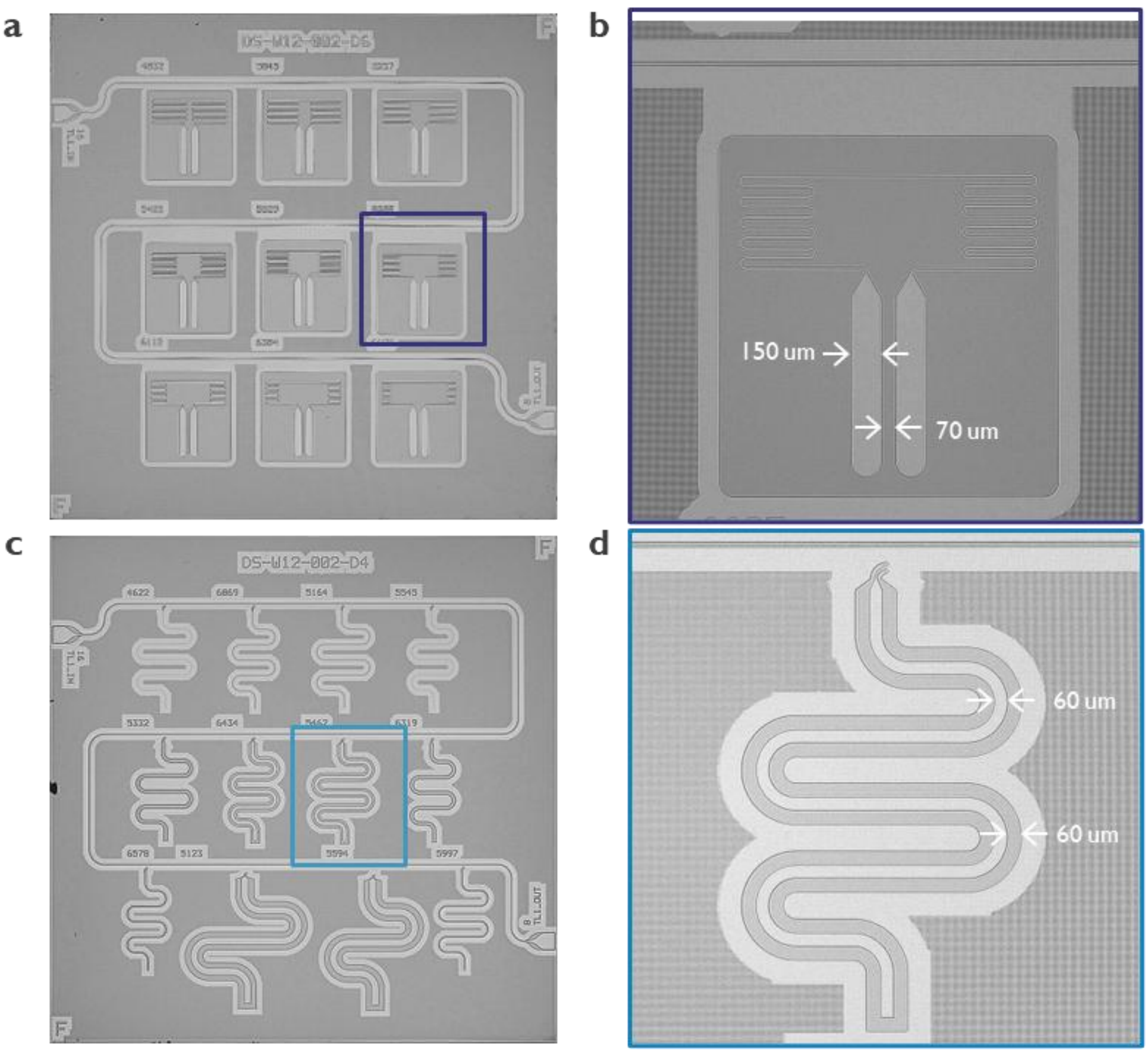

*Figure S5: Optical micrograph of fabricated resonator samples. **a** lumped element resonator sample with **b** zoom-in on a single lumped element resonator (LE70) inductively coupled the feedline. The capacitor pads are 150 µm wide and separated by 70 µm (LER70) **c** coplanar waveguide resonator sample) with **d** zoom-in on resonator (W60) capacitively coupled to the feedline.*

Resonator samples are wire-bonded directly to the SMP bulkhead connector and to the aluminium package using 25 μm diameter Al wire bonds. Wire-bond lengths are minimized to reduce impedance mismatch arising from parasitic inductances. A micrograph of a bonded sample is shown in Figure S6.

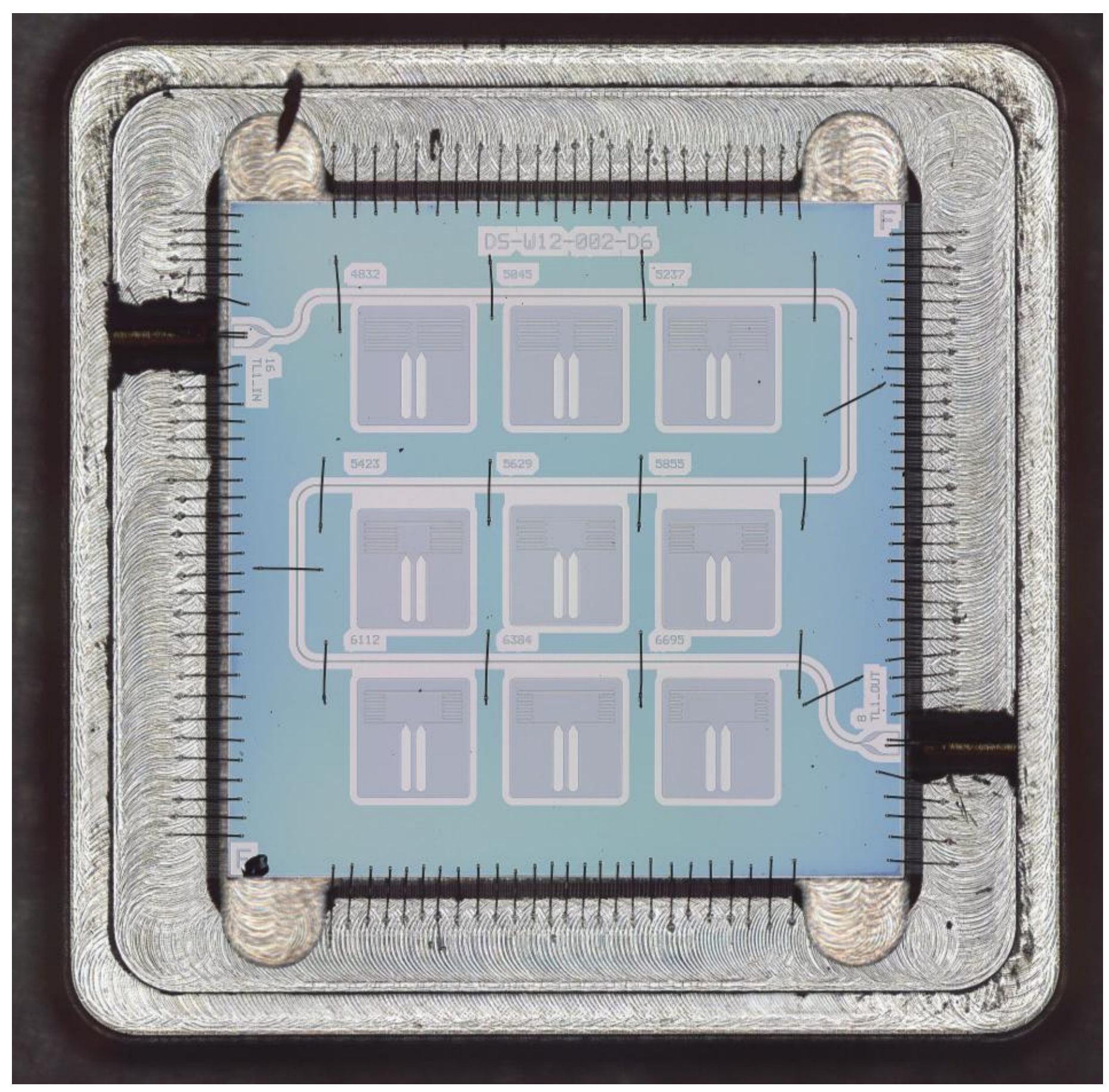

*Figure S6: Wire-bonded sample to an Al package. Representative image of a LE70 resonator chip wirebonded to the SMP bulkhead (signal) and Al package (ground).*

## Impact of metal-substrate interface on resonator loss

In this section we comparative $Q_{TLS}$ measured for RT Ta (with 10 nm-thick Nb liner, filled triangles) and hot Ta[27,34] (400 °C deposition without a liner, filled stars), with and without HF post-treatment (pink/violet vs. grey) (Figure S7). Measurements are performed on coplanar waveguide resonators fabricated using an earlier device geometry, with a central conductor width of 24 μm and a gap of 12 μm (W24S12). This chip contains multiple resonators of the same design, enabling improved statistical comparison of the different processing conditions. Coplanar waveguide resonators with W24S12 dimensions are also more sensitive to interface losses than LE resonators, making them suitable for evaluating the metal–substrate interface. The substrate for these measurements is REF_3k; however, the choice of substrate is not critical for the observed trends, as the loss is dominated by surface contributions. In the absence of HF treatment, both RT Ta and hot Ta exhibit a median $Q_{TLS}$ of approximately $2 \times 10^6$.

Following HF post-treatment, the median $Q_{TLS}$ for RT Ta increases to ~$9 \times 10^6$, whereas hot Ta reaches only ~$4 \times 10^6$. This twofold improvement for RT Ta indicates an enhanced metal-substrate (MS) interface quality provided by the 10 nm Nb liner. Notably, prior studies on RT Ta reported no significant improvement in $Q$ factors relative to hot Ta with 5 nm thick Nb liner.[37] In contrast, our results demonstrate a substantial enhancement, which may be linked to the increased Nb liner thickness of 10 nm.

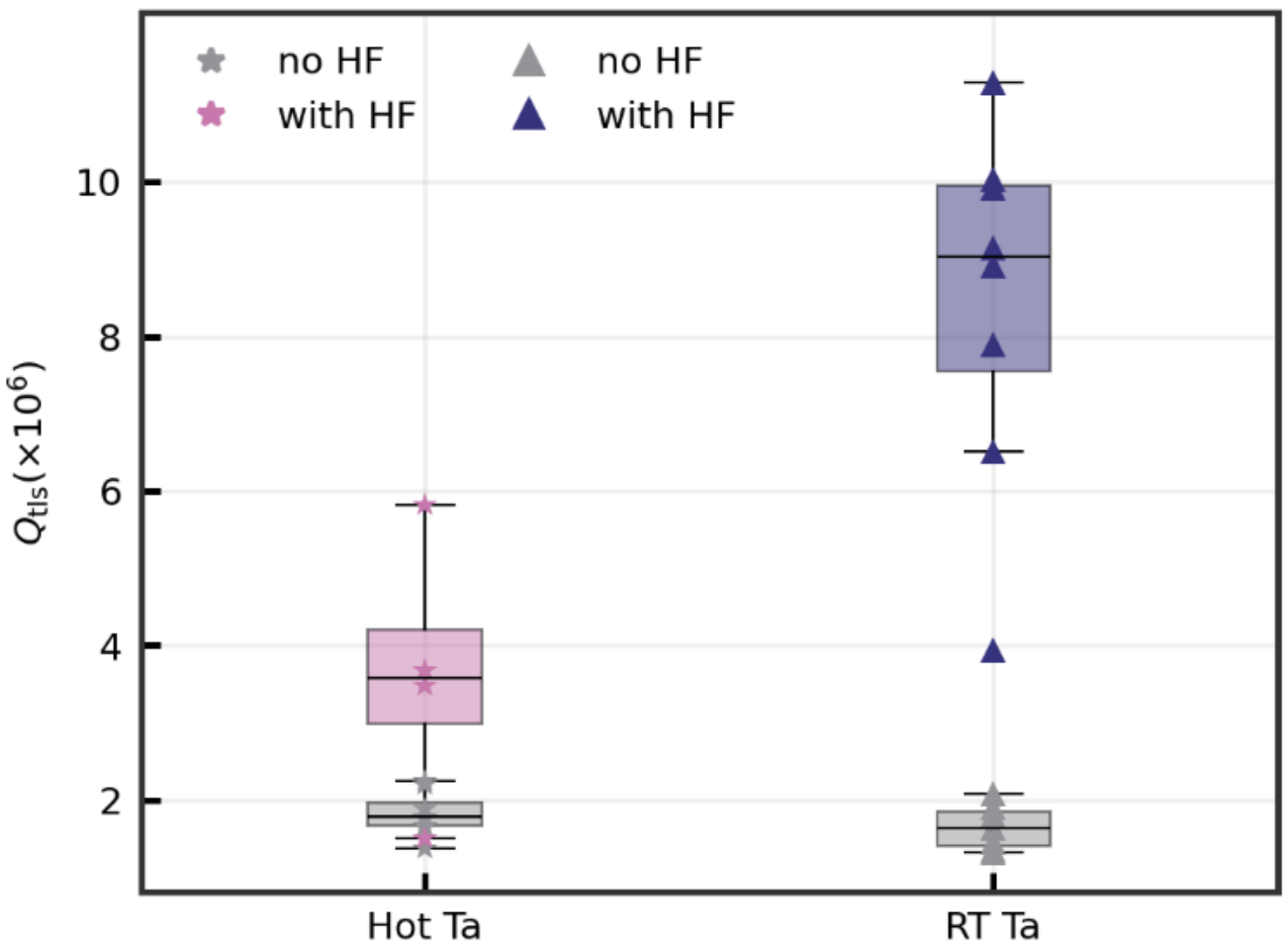

*Figure S7: $Q_{TLS}$ comparison between RT Ta (filled triangles) and hot Ta[27,34] (filled stars) measured on coplanar waveguide resonators with central conductor width of 24 μm and gap between central conductor and ground plane of 12 μm, with and without HF post-treatment (pink/violet vs. grey). RT Ta with Nb liner gives a two-fold increase in resonator Q-factor with a HF post-treatment. The data is taken from samples with the REF_3k substrate.*

To investigate the potential influence of Nb liner thickness on superconducting properties, cryogenic DC transport measurements were performed on RT Ta films using a test pattern with a length to width ratio (number of squares) of ~602 (Figure S8). The critical field was measured as a function of temperature using a Kiutra adiabatic demagnetization refrigerator, until the sample transitioned to the normal state. A comparison of the extracted parameters with previously reported values[37] is provided in Supplementary Table 1. While the films in this study exhibit a slightly higher residual resistivity ratio (RRR), most properties are comparable to prior work.[37] The notable exception is the superconducting critical temperature ($T_c \approx$ 4.9 K), which exceeds the value of ~4.3 K typically reported for both RT Ta and hot Ta. This enhancement in $T_c$ can be likely attributed to a proximity effect induced by the Nb liner. Further studies are needed to clarify the influence of liner thickness on resonator quality factors.

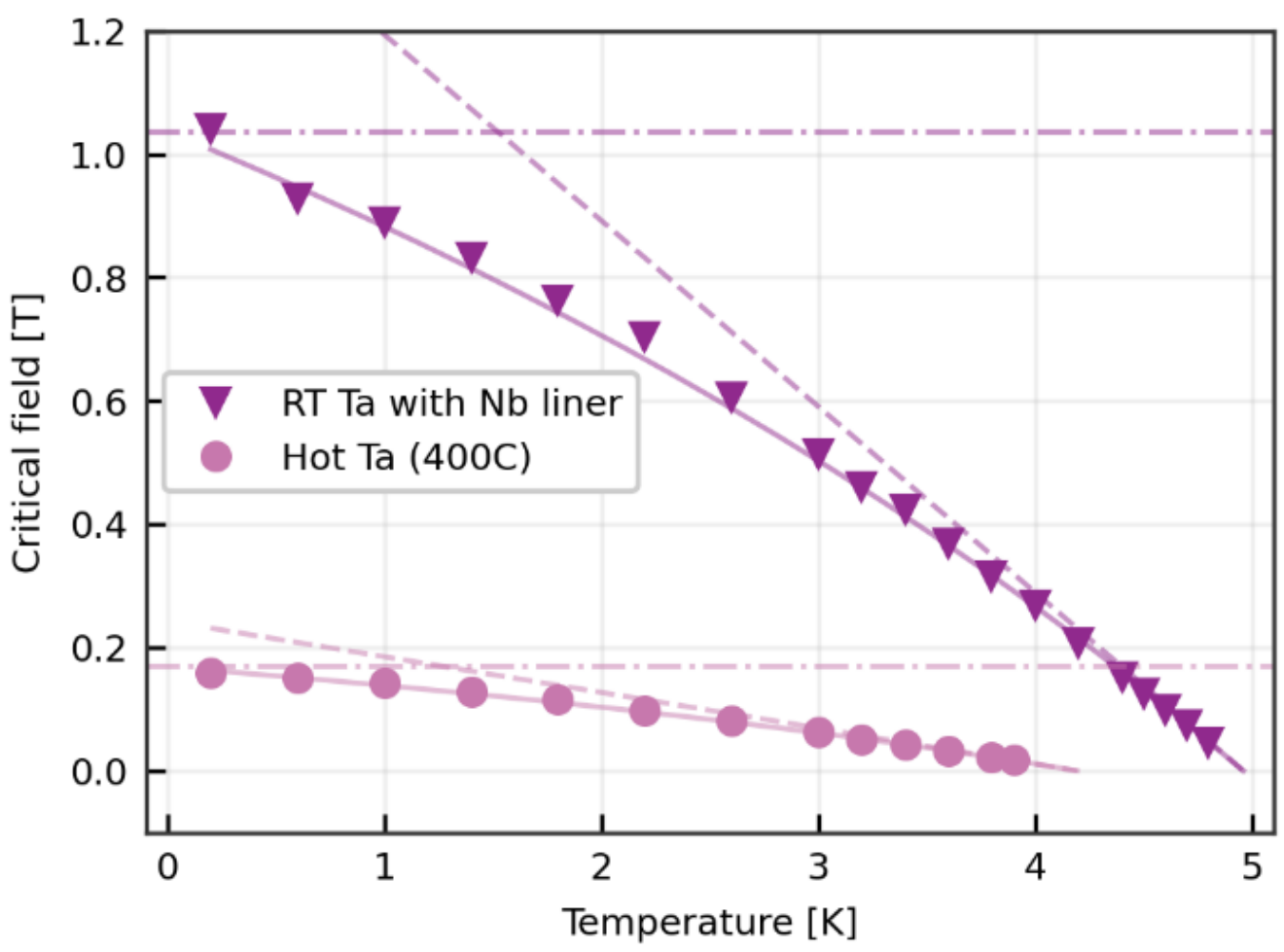

*Figure S8: Phase diagram from transport measurements on RT Ta (filled triangles) and hot Ta (filled circles). The data is fit to a model that considers a linear slope close to $T_c$ (dashed) and saturation (dash-dot) at temperatures $T \ll T_c$.*

*Supplementary Table1: Electrical and superconducting properties of RT Ta and hot Ta films. Parameter extraction follows the same methodology and formulae as in Refs. [37,45,46].*

| Parameter | RT Ta (this work) | RT Ta Ref. [37] | hot Ta (this work) | hot Ta Ref. [37] |
|---|---|---|---|---|
| Critical temperature, $T_c(\mathrm{K})$ | 4.96 | 4.45 | 4.20 | 4.30 |
| Sheet resistance @300K, $R_{300K}(\Omega/\square)$ | 1.91 | 2.94 | 1.72 | 1.57 |
| Sheet resistance @5K, $R_{5K}(\Omega/\square)$ | 0.49 | 1.16 | 0.03 | 0.08 |
| Residual resistance ratio, $RRR$ | 3.9 | 2.6 | 62.7 | 20.4 |
| Sheet kinetic inductance, $L_k(fH/\square)$ | 136 | 360 | 9 | 25 |
| Slope (from fit), $-dB_c^2/dT_c(\mathrm{mT/K})$ | 301 | 372 | 58 | 47 |
| Mean-free path, $l\ (nm)$ | 4.5 | 3.7 | 24 | 29 |
| London penetration depth, $\lambda_L(nm)$ | 109 | 173 | 25 | 44 |
| Critical field, $B_{c2}(0)\sim 0.693T_c \times \left.\frac{dB_{c2}}{dT}\right\|_{T=T_c}$ (T) | 1.03 | 1.15 | 0.17 | 0.14 |
| Critical field, $B_{c2} = T_c \times \left.\frac{dB_{c2}}{dT}\right\|_{T=T_c}$ (T) | 1.49 | 1.6 | 0.24 | 0.2 |
| BCS coherence length, $\xi_0^{BCS}(nm)$ | 15 | 14 | 37 | 40 |
| GL coherence length, $\xi_0^{GL}(nm)$ | 18 | 14 | 44 | 40 |

## Energy participation ratio scaling

Energy participation ratios (EPRs) for dielectric regions in CPW and LE resonator designs are computed using finite-element simulator Ansys Maxwell. For CPW resonators, the model includes a cross-section of a center trace and a ground plane, while for LE resonators, a cross-section of the dipole capacitor pads. The simulations assume dielectric constants of 11.9 for the Si substrate, 4 for the substrate-air (SA) interface, and 10 for both the metal-substrate (MS) and metal-air (MA) interfaces.[25] All interfaces are modelled with equal thickness, swept between 6 and 20 nm, and subsequently rescaled to an effective thickness of 3 nm to allow comparison with reported loss tangents.[29] A 100 nm silicon recess is included. The resulting EPRs are summarized in Supplementary Table 2. While W120 data is not being used, we keep it here for completeness. To independently validate the simulated participation ratios, selected structures were additionally simulated using the finite-element solver Palace with adaptive mesh refinement and explicitly meshed MS, MA and SA interface layers of 10 nm thickness. Participation ratios were subsequently rescaled to an effective oxide thickness of 3 nm. The extracted participation ratios agree within 3–9% with the values obtained from the Ansys Maxwell simulations used throughout this work, with convergence better than 0.1% during the final refinement steps, supporting the robustness of the extracted trends and fitted loss tangents.

*Supplementary Table 2: Energy participation ratios obtained from Ansys Maxwell 2D finite-element simulations.*

| | LE70 | W120 | W60 | W30 | W10 | W3 | W1 |
|---|---|---|---|---|---|---|---|
| MS ($\times 10^{-4}$) | 0.98 | 0.88 | 1.59 | 2.88 | 7.29 | 19.43 | 45.09 |
| SA ($\times 10^{-4}$) | 0.37 | 0.33 | 0.60 | 1.12 | 2.99 | 8.59 | 22.03 |
| MA ($\times 10^{-6}$) | 1.86 | 1.62 | 3.10 | 6.00 | 17.14 | 54.29 | 158.25 |
| Substrate | 0.922 | 0.922 | 0.921 | 0.920 | 0.917 | 0.907 | 0.882 |
| MS/MS [W30] | 0.341 | 0.306 | 0.551 | 1.0 | 2.533 | 6.754 | 15.674 |
| SA/SA [W30] | 0.329 | 0.292 | 0.538 | 1.0 | 2.660 | 7.642 | 19.600 |
| MA/MA [W30] | 0.309 | 0.270 | 0.516 | 1.0 | 2.856 | 9.042 | 26.357 |

Supplementary Table 2 further shows that the normalized (to W30) participation ratios for all three interfaces (MS, SA, and MA) exhibit similar scaling trends for larger geometries (LE70-W10). However, this correspondence breaks down for smaller feature sizes (W3-W1), where the metal–air (MA) participation increases significantly more rapidly than both MS and SA. This highlights the stronger geometric dependence of the MA interface in the small-feature regime. The stronger scaling of the MA participation for smaller geometries is consistent with enhanced electric field localization at metal edges, which increases the relative contribution of edge-associated dielectric regions as the feature size is reduced. However, given the sensitivity of surface participation to field singularities and meshing near conductor edges, we cannot fully exclude a contribution from numerical effects in the simulation.

## Anchor analysis

The EPR analysis presented in the main text employs an effective two-component model (Eq. 1), with the metal-substrate participation ratio used as the effective surface participation axis. Since multiple interface regions contribute to the total electric-field energy, this choice requires justification. To assess its uniqueness, we perform an anchor analysis comparing the consequences of representing interface loss through either the metal-substrate (MS) or metal-air (MA) participation channels.

In this analysis, W1 serves as the anchor resonator. A prescribed fraction of the measured TLS loss in W1 is assigned to a selected interface channel, and the corresponding interface loss tangent is obtained by dividing this loss contribution by the simulated participation ratio of that channel.

$$\tan\delta_{\mathrm{anchor}} = \frac{1}{p_{\mathrm{anchor}}^{W1}} \times \frac{f}{Q_{\mathrm{TLS}}^{W1}},$$

here $f$ is the prescribed anchor fraction, $Q_{TLS}^{W1}$ is the measured quality factor of W1, and $p_{\mathrm{anchor}}^{\mathrm{W1}}$ is the participation ratio of the selected interface channel in W1.

While the anchored loss tangent of one interface is fixed, the complementary interface loss tangent is determined from a fit to the dataset using the following equation:

$$\frac{1}{Q_{\mathrm{TLS}}} = p_{\mathrm{MA}} \tan\delta_{\mathrm{MA}} + p_{\mathrm{MS}} \tan\delta_{\mathrm{MS}} .$$

This procedure is repeated for both MA- and MS-anchored fits. Representative fits and the resulting parameter evolution are shown in Figure S9.

The anchor analysis demonstrates that the MA and MS participation ratios are not orthogonal within the present resonator family. Over a broad range of anchor assumptions, increases in one interface contribution can be compensated by decreases in the other while maintaining acceptable agreement with the measured participation-ratio dependence. Consequently, the dataset does not uniquely determine how interface-related loss should be partitioned between MA and MS channels.

Despite this non-orthogonality, a pronounced asymmetry is observed. When the MA contribution is increased, the fitted MS loss tangent decreases only gradually and remains finite throughout the entire anchor sweep, remaining of order $10^{-4}$ even when the anchor resonator loss is fully attributed to the MA channel. In contrast, when the MS contribution is increased, the fitted MA loss tangent rapidly decreases and eventually collapses to the lower fitting bound. Beyond approximately 80–90% assigned MS contribution, the measured participation-ratio dependence can be reproduced without requiring a significant MA contribution.

These results indicate that the observed scaling is more naturally represented using an MS-based participation axis. While the present analysis does not uniquely separate MA and MS dissipation channels, it demonstrates that a finite MS contribution remains preferred throughout the MA-anchored analysis, whereas substantial MA dissipation is not required to reproduce the measured participation-ratio dependence. Accordingly, the effective loss tangent reported in the main text should be interpreted as an effective interface-loss parameter projected onto the MS participation axis, rather than as a uniquely determined microscopic MS-interface loss tangent.

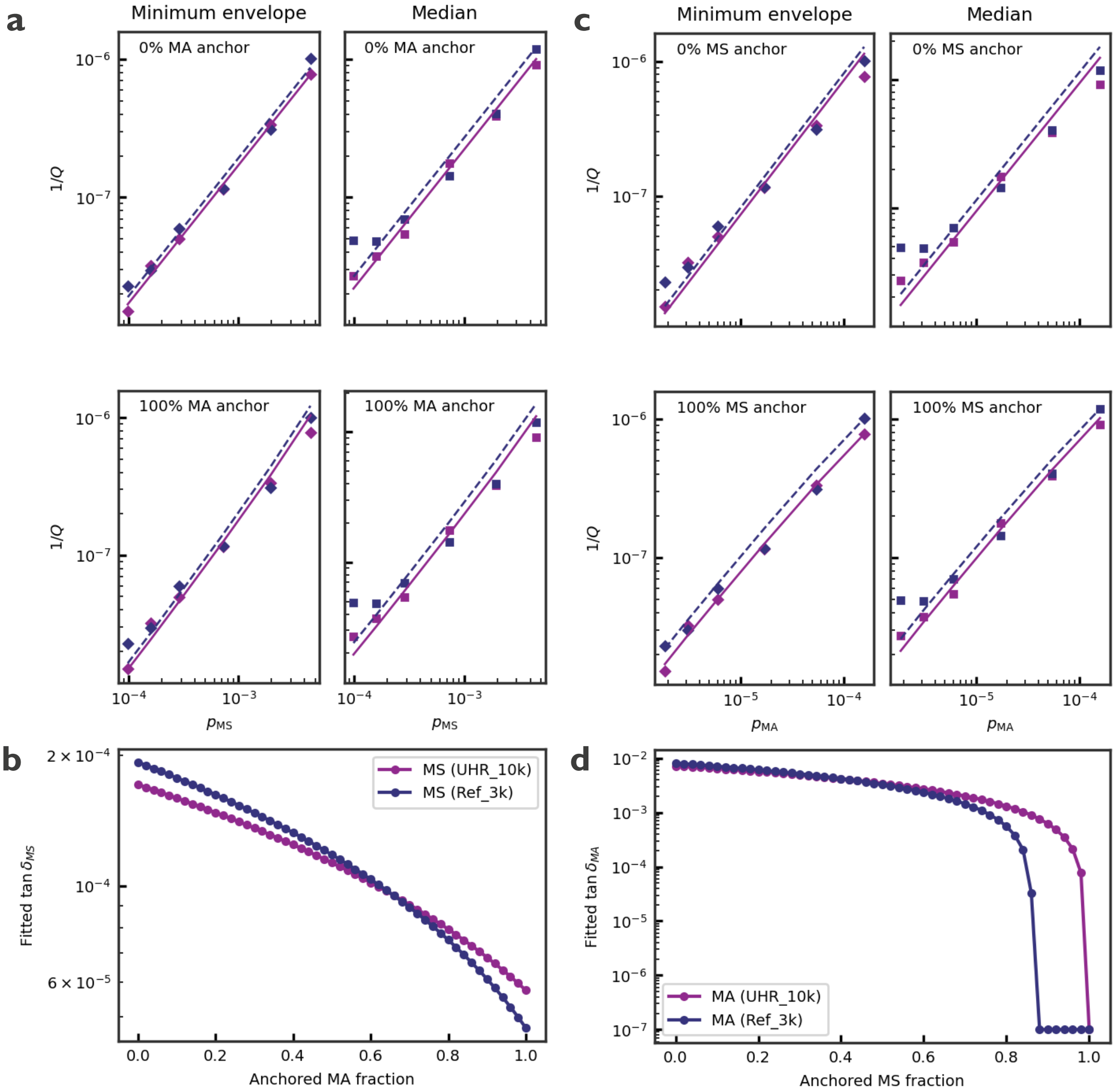

*Figure S9: Anchor analysis of interface participation assignment.* ***a.*** *Representative MA-anchored fits for the minimum-loss and median-loss analyses. The MA contribution is constrained to account for a prescribed fraction of the loss in an anchor resonator, while the corresponding MS loss tangent is fitted.* ***b.*** *Evolution of the fitted MS loss tangent as the imposed MA contribution is increased. The extracted MS loss tangent decreases gradually but remains finite throughout the anchor sweep.* ***c.*** *Representative MS-anchored fits for the minimum-loss and median-loss analyses, where the MS contribution is fixed and the MA loss tangent is fitted.* ***d.*** *Evolution of the fitted MA loss tangent as the imposed MS contribution is increased. In contrast to (b), the fitted MA loss tangent decreases rapidly and eventually collapse to the lower fitting bound. The similar behaviour observed for both minimum-loss and median-loss analyses demonstrates that the conclusions are not sensitive to the metric employed. While the anchor analysis reveals substantial non-orthogonality between MA and MS participation ratios, the observed asymmetry indicates that the measured scaling is more naturally represented using an MS-based participation axis, motivating its use as the effective surface participation axis in the main text.*

## Statistical analysis of variance (ANOVA)

ANOVA (analysis of variance) and Tukey's HSD post-hoc tests were employed to determine statistical significance between substrate groups. Family-wise error rate (FWER) was controlled to minimise false positives, with p-values $<0.05$ indicating significant differences.

A one-way ANOVA (analysis of variance) was conducted to compare the low-power internal Q-factors measured across four groups: REF_3k, UHR_10k, and UHR_11k. The analysis revealed a significant effect of group, $F = 7.79$, $p = 0.0014$, where the F-statistic represents the ratio of variance between groups to variance within groups.

Post hoc comparisons using Tukey's HSD test with a family-wise error rate (FWER) of 0.02 indicated that UHR_10k was significantly different from both UHR_11k and REF_3k. In contrast, UHR_11k and REF_3k were not significantly different from one another. The adjusted p-values were 0.0022 for REF_3k vs UHR_10k, 0.0166 for UHR_10k vs UHR_11k, and 0.993 for REF_3k vs UHR_11k.

The group means (± standard deviation) were as follows: REF_3k = 22.2 ± 10.0 million, UHR_11k = 22.8 ± 8.2 million, UHR_10k = 38.07 ± 16.62 million. The effect size for the ANOVA was $\eta^2 = 0.275$, indicating that approximately 27.5% of the variance in the Q-factors can be attributed to differences between groups, representing a large effect size. This suggests that substrate group has a substantial influence on the measured Q-factor.

Prior to conducting ANOVA, assumptions were verified. Normality was assessed using the Shapiro-Wilk test for each group, and no significant deviations from normality were observed ($p > 0.5$ for all groups). Homogeneity of variances was assessed using Levene's test. The test did not indicate a significant violation of the equal-variance assumption ($p = 0.055$), although the result was close to the significance threshold. These results support the appropriateness of using one-way ANOVA and Tukey's HSD for post hoc comparisons and justify the use of means as the measure of central tendency.